\documentclass[%
 reprint,
 amsmath,amssymb,
 aps,prl,
floatfix
]{revtex4-2}
\usepackage[utf8]{inputenc}
\UseRawInputEncoding
\usepackage{xcolor}
\usepackage{graphicx}
\usepackage{dcolumn}
\usepackage{bm}
\usepackage{subfig}
\usepackage{float}
\usepackage{hyperref}

\usepackage{caption}
\begin{document}

\title{Role of bubble positioning in force induced melting of DNA}

\author{Bidisha Mukherjee$^{1}$, Amit Raj Singh$^{2}$ and Garima Mishra$^{1,*}$, }

\affiliation{${^1}$Department of Physics, Ashoka University, Sonipat 131029, India}
\affiliation{${^2}$Department of Physics, Graphic Era Hill University, Dehradun, India}

\date{\today}

\begin{abstract}
  
We investigate the role of bubble positioning in the 
force-induced melting of double-stranded DNA using two 
distinct approaches: Brownian Dynamics simulations and 
the Gaussian Network Model. We isolate the effect of 
bubble positioning by using DNA molecules with 
$50\%$ AT - $50\%$ GC base-pair composition which 
ensures constant enthalpy. Our results reveal that 
it is not just the sequence itself, but its specific 
arrangement that influences DNA stability. 
We examine two types of DNA sequences containing 
a block of either AT or GC base-pairs, resulting in 
the formation of a large bubble or a smaller bubble 
within the DNA, respectively. By systematically 
shifting these blocks along the strand, we investigate 
how their positioning influences the 
\texorpdfstring{force-temperature}{force-temperature} 
phase diagram of DNA. Our Brownian dynamics simulations 
reveal that, at high forces, melting of the entire DNA 
strand is initiated after stretching $\approx 9$ 
GC base pairs, independent of the specific base-pair 
sequence. In contrast, no such characteristic length 
scale is observed in the Gaussian network model. 
Our study suggests that free strand entropy plays a 
significant role in determining the 
\texorpdfstring{force-temperature}{force-temperature} 
phase diagram of the DNA.
  
  \end{abstract}

\maketitle
\section{Introduction}

The process of DNA strands separation (double stranded DNA 
to single stranded DNA) plays a fundamental role in replication 
\cite{kornberg1984dna} and transcription, making it a widely 
studied topic in molecular biophysics \cite{Mol-Bio}. In 
traditional view of temperature-induced melting, AT rich 
regions of the DNA are more prone to thermal destabilization 
due to lower stability of AT base-pairs compared to GC base-pairs, 
reflecting the inherent energetic heterogeneity of DNA base-pairing. 
This differential stability of base-pairs leads to the formation 
of localized bubbles of unpaired bases. These bubbles can be 
utilized by regulatory proteins or functional enzymes to carry 
out their roles \cite{Park-Roberts2006,tapia2014mesoscopic}. 
Recent studies suggest an increased likelihood of large bubbles 
forming at functionally important sites, such as gene promoter 
region \cite{Choi_gkh335,G.Kalosakas_2004,van2005can,Hillebrand}. 
These bubbles, which can span multiple base-pairs, expand as 
the temperature rises, eventually resulting in complete melting 
of the DNA strand. The differential melting curves of DNA show 
length dependent behavior: the melting curve shows single or 
multiple peaks for a very short ($10^2 - 10^3$ bp) or 
long ($10^3 - 10^4$ bp) sequence of DNA \cite{VOLOGODSKII20181,Sergei1984}. 
A very long chain of DNA ($10^6$) again results only in a 
single broad peak in the melting curve \cite{WARTELL198567}. 
The experimental melting curve of DNA for a broad range of 
sequence length has been reproduced by re-parameterizing the 
loop weight contribution \cite{Blossey2003} embedded in DNA chain. 

A fair amount of understanding is available for temperature 
induced melting of DNA \cite{mergny2003analysis,VOLOGODSKII20181}. 
The modern view of strand separation \textit{in vivo}, however, 
is by a more mechanical procedure of force or torque induced 
unzipping and supersedes the traditional view of melting 
because of the lack of extreme temperature environment 
inside the cell for the melting of DNA ($70 - 100^o$C) 
\cite{bhattacharjee2000unzipping}. Single-molecule force 
spectroscopy (SMFS) experiments contribute significantly 
to the understanding of force induced melting behavior of 
DNA \cite{ritort2006single,bustamante2000single}. A first 
experimental step towards understanding the interplay 
between thermal and force-induced separation of DNA was 
taken by Danilowicz 
{\it et al} \cite{SMF}. This study showed that the force 
required to unzip the DNA decreases with increasing 
temperature and is consistent with theoretical predictions 
\cite{bhattacharjee2000unzipping}.
 
 The pre melting of AT rich region, which results in multiple 
 peaks in the melting profile of DNA, may also affect the force 
 induced behavior of DNA at intermediate length scales 
 \cite{singh2023pulling,bubbleweight2003}. Models that explicitly 
 incorporate thermally generated bubbles 
 \cite{Theodrakopoulos,peyrard2004nonlinear} have been shown 
 to provide a more accurate description of force-induced DNA 
 melting compared to those that neglect bubble formation 
 \cite{dauxois1993entropy}. The role of explicit bubbles, 
 particularly in the form of DNA hairpins, has been investigated 
 in the context of force-induced DNA melting, and corresponding 
 \texorpdfstring{force-temperature}{force-temperature} phase 
 diagrams have been explored \cite{MishraJCP,rudra2023force,chauhan2022can}. 
 However, the influence of sequence heterogeneity in 
 double-stranded DNA can result in implicit bubbles, 
 depending on the positioning of AT-rich regions. 
 These bubbles do not form well-defined secondary structures 
 like hairpins but can still influence the mechanical response 
 of DNA to external force. The influence of such implicit 
 bubbles on the \texorpdfstring{force-temperature}{force-temperature}
 phase diagram of DNA remains elusive, owing to significant 
 variability in their size and location along the DNA strand. 
 This aspect warrants further investigation.

An interesting observation from single-molecule force spectroscopy 
(SMFS) experiments is that applying force along the helical axis 
results in sequential stretching of DNA bases \cite{essevaz1997mechanical}, 
with the shearing force saturating at a finite value in the long-length 
limit \cite{hatch2008demonstration,kuhner2007force}. These 
experimental findings are consistent with earlier theoretical predictions 
\cite{de2001maximum, mishra2011effect,prakash2011shear,Chakraborty}. 
Other studies also examined the effects of applying force 
perpendicular to the helical axis, revealing a threshold length 
beyond which the applied force ceases to affect the base-pairs 
\cite{singh2016opening,singh2015pulling}. The earlier studies 
on short DNA chain put constraint on the effect of any such 
length scale in DNA unzipping (where force is applied in 
the direction perpendicular to the helical axis) 
\cite{Gonzalez_2009,Mishrajcp2013}. It also restricts the 
size and positioning of bubble, which may otherwise be 
recruited to a location along the DNA depending on the AT-rich sequence.  

The focus of the current study is to understand the role of 
energetic heterogeneity which leads to implicit bubble of 
varying size, at different locations along the DNA, on 
force-temperature diagram. To explore this, we examine 
relatively longer DNA sequences than those used in earlier 
studies \cite{Gonzalez_2009,MishraJCP}, allowing us to 
investigate the role of a critical length scale in 
force-induced unzipping \cite{singh2016opening} if 
any. In Sec. II, we present the model along with the 
specifics of the Brownian dynamics (BD) simulations 
and the Gaussian network model (GNM). Section III 
discusses the formation of bubble and its role on 
force-temperature diagram ($f-T$),
and compares results obtained from BD simulations 
with those from the GNM. A summary of our findings 
is provided in Sec. IV. Additional details are 
available in the Supplemental Material [..], including several figures.
\section{Models and Methods}
\subsection{Off lattice coarse grained modeling of DNA}

In this study, we adopt a minimal off-lattice coarse-grained DNA model, 
where a single bead represents one nucleotide. This model 
captures the energetic heterogeneity arising from AT and 
GC base pairs as well as the excluded volume effects. 
The model potential energy considered here is 
\begin{eqnarray}
  \begin{array}{ll}
  
 \end{array}& E = &  \sum_{i=1}^{N-1}K(d_{i,i+1}-d_0)^2 + 
 4 \sum_{\rm Non-native}\bigg(\frac{\sigma_{i,j}}{d_{i,j}}\bigg)^{12}\nonumber\\
 & &+4{\epsilon^{HB}_{i,j}}\sum_{\rm Native-contacts} \bigg[\bigg(\frac{\sigma_{i,j}}{d_{i,j}}\bigg)^{12}  -\bigg(\frac{\sigma_{i,j}}{d_{i,j}}\bigg)^6\bigg], \label{eq:1}
 \end{eqnarray}

with $N (= 128)$ represents the total number of beads in the model system. 
The first term in the potential energy function  represents the 
harmonic potential, where $K$ denotes the spring constant. 
In this model, $K$ is set to $100$ to simulate covalent bonds 
between consecutive beads. The distance $d_{i,j}$ between 
beads $i$ and $j$ is defined as $|\vec{r}_i - \vec{r}_j|$, 
where $\vec{r}_i$ and $\vec{r}_j$ denote the respective 
positions of beads $i$ and $j$, with $i$ and $j$ ranging 
from $1$ to $N$. Distances are dimensionless, with $\sigma_{i,j} = 1$, 
and the equilibrium bond distance is $d_0 = 1.12$. 
The energy parameters in the potential energy function 
are expressed in units of $k_B T$, where $k_B$ is the 
Boltzmann constant, and $T$ is the temperature. The second 
term in the potential energy introduces a repulsive 
potential to prevent overlap between non-native pairs of 
monomers in the chain \cite{MishraJCP}. The third term accounts 
for the hydrogen bonding between base-pairs, incorporating 
$\epsilon^{HB}_{i,j}$, which facilitates pairing between 
non-bonded monomers, $i$ and $N-i+1$. The interaction energy 
$\epsilon^{HB}_{i,j}$ is assigned values to 
capture the relative strengths of base-pairing between AT and 
GC pairs ($\epsilon_{i,j}^{AT} = 0.67$ and $\epsilon_{i,j}^{GC} = 1.0$), 
maintaining  ratio of $1.5$.

We obtain the dynamics of the DNA by using the following Brownian 
equation \cite{kumar2010biomolecules}
\begin{equation}
    \zeta \frac{d\mathbf{r_i}}{dt} = \mathbf{F_c} + \Gamma.
\end{equation}

Here $\zeta(= 50)$ is the friction coefficient. 
$\mathbf {F_c = -\nabla E}$ is the conservative force and the random force 
$\Gamma$ is a white noise, with zero mean and correlation 
$<\Gamma(t)\Gamma(t')>= 2\zeta k_BT\delta_{i,j} \delta(t-t')$. 
To study the unzipping of DNA subjected to the external force 
$f$ ($f$ is applied to one end of the DNA chain, called the 
force end or free end, and the other end is connected by 
covalent bond, referred to as the bound end),
we add an energy $-fy$ to the total energy of 
the system given by Eq.1, $y$ is the distance between 
the ends where force is applied. These equations of motion are 
integrated {\it via} the Euler method with time step $\delta t =0.0015$ 
for $10^8$ iterations at different $T$ and $f$. 

We obtain the $f-T$ diagram by calculating the specific heat ($C_v$) 
as a function of $T$ at a given $f$. The peak in the specific heat 
corresponds to the melting temperature ($T_m$) at given 
$f$ \cite{mishra2011effect,Li-Suan1999}. Another widely used 
definition for $T_m$ is a temperature at which half of the 
base-pairs of the DNA chain gets open \cite{sengupta2024large,mishra2024sheetlike}. 
By using the above two definitions, we obtained a cut-off distance 
for base-pairing, above which base-pairs are considered to be open.

 \subsection{Gaussian Network Model}

The Gaussian Network Model (GNM), originally developed for studying 
protein dynamics, has been extended to model DNA and is referred to 
as the Gaussian Bond Elasticity (GBE) model \cite{Singh16, SinghForce}. 
This approach integrates the elastic framework of GNM in three-dimensional 
(3D) space with bond binding energies to capture the structural 
and energetic features of DNA. In this coarse-grained network model, 
each nucleotide is represented by three nodes, and all key 
interactions, covalent bonds, hydrogen bonding, and base stacking 
are modeled using harmonic potentials. The Hamiltonian of the system 
is expressed as a sum of these harmonic terms, with spring constants 
derived from quantum mechanical calculations of the respective 
interaction energies. In the presence of a pulling force 
applied perpendicularly at one end of
DNA, an elastic energy contribution is added to the Hamiltonian.
The GBE allows for efficient sampling of the equilibrium behavior 
of DNA in terms of the normal modes of the network connectivity matrix 
which in turn allows to calculate the partition function and Helmholtz free energy.
The most probable denaturation pathway (MPDP) is identified by 
evaluating the Helmholtz free energy associated with intermediate 
denaturation states (as given in 
Eq. 9 of \cite{SinghForce}) and applying a steepest descent-like 
algorithm \cite{Singh16, SinghForce}. This model has been proven 
to capture the effect of loop entropy and the existence of 
intermediate states for DNA molecules very accurately and is 
computationally cheap \cite{Singh16,SinghForce,chauhan2022can,rudra2023force} 
in comparison to the Brownian dynamics simulations. We employ the 
same procedure as described in \cite{SinghForce} to determine the 
force-temperature (\texorpdfstring{$f-T$}{f-T}) phase diagram. 
For a fixed force $f$, we identify the temperature $T_m$ at which 
half of the base pairs in the DNA chain are denatured.

\begin{figure}[t]
\centering
\includegraphics[width=\linewidth,scale=0.125,trim={0.5in 0.2in 0.1in 0.1in},clip]{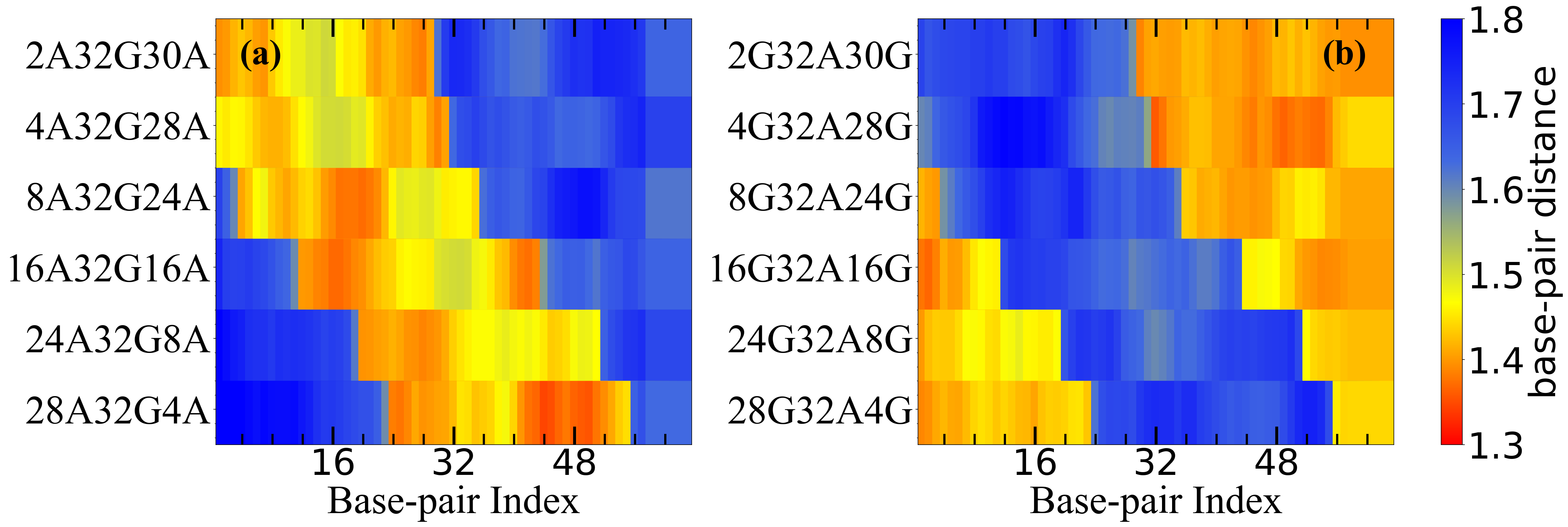}
\caption{The color map shows the variation of average base-pair 
distance along the DNA strand at $f=0$ and at $T=0.13$ (below 
the $T_m$), using BD simulation, for Sequences with (a) moving GC 
block, (b) moving AT block. The x-axis represents the base-pair 
index, while the y-axis lists different DNA sequences.}
  \label{bubbbleplot}
\end{figure}

 To assess the impact of implicit bubble resulting from sequence 
 heterogeneity and it's positioning on $f-T$ diagram, we consider 
 specifically designed DNA sequences of $64$ base-pairs, 
 with a $50\%$ GC and $50\%$ AT base-pair which ensures constant 
 enthalpy. We explore two distinct types of sequence distributions. 
 In the first distribution, we considered sequences where a 32G block 
 is located at different positions along the DNA, with a surrounding 
 distributed AT region. This type of distribution can generate 
 implicit bubbles of varying sizes at the bound ends, depending on 
 the specific sequence. The sequences considered in our study are - 
 seq1: 2A32G30A, seq2: 4A32G28A, seq3: 8A32G24A, seq4: 16A32G16A, 
 seq5: 24A32G8A, seq6: 28A32G4A. Here, we are mentioning the sequence 
 of one strand. Needless to say, the sequence of the other strand is 
 complementary to the first one. For the second type of distribution, 
 the 32G block is dispersed along the DNA, with a localized 32A block, 
 leading to configuration that features an implicit bubble of fixed 
 size at different positions along the strand. Examples include seq1: 
 2G32A30G, seq2: 4G32A28G, seq3: 8G32A24G, seq4: 16G32A16G, Seq5: 
 24G32A8G, seq6: 28G32A4G.

\section{Results}

\begin{figure}[t]
 \centering 
\includegraphics[width=\linewidth,scale=0.20,trim={0.00in 3.0in 0.0in 0.3in},clip]{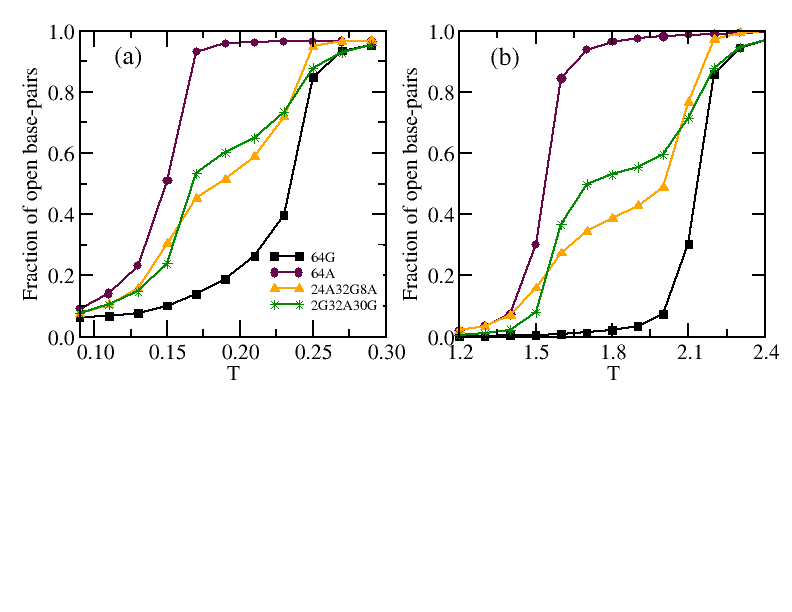}
  \caption{ The variation of fraction of open base pairs as a function of 
  $T$ at $f=0$ from (a) BD simulation and (b) GNM for different sequences.}
  \label{contactplot2}
\end{figure}

 To ensure that the model accurately captures the presence of weak and 
 strong energetics of base-pairs, we analyze the base-pair distances between 
 native pairs at $f=0$ and $T=0.13$. We observe how base-pair distances vary 
 across different sequences, with respect to the base-pair index using a 
 color-coded scheme (Figure \ref{bubbbleplot}). The weakly interacting (AT) 
 base-pairs, exhibit larger distances (indicated in blue) compared to the 
 strongly interacting (GC) base-pairs, which are represented in yellow colors. 
 AT/GC base-pairs located at the free ends of the DNA display larger distances 
 due to the additional contribution of end entropy, in contrast to AT/GC pairs 
 near the bound end of the DNA. Furthermore, AT base-pairs within the 
 chain show bubble formation with varying sizes at loop end (Figure \ref{bubbbleplot}a) 
 and at different locations (Figure \ref{bubbbleplot}b), as seen in (Figure \ref{bubbbleplot}). These observations support the conclusion that our model effectively captures the 
 energetics of AT/GC base-pairs in DNA. By applying a cutoff distance 
 of $1.5$ \cite{sengupta2024large,mishra2024sheetlike}, base-pairs with 
 separations exceeding this cutoff distance are classified as open. 
 This enables us to quantify the fraction of open base-pairs in the 
 DNA and, consequently, determine the melting temperature $T_m$ at 
 which half of the base-pairs get open.

We observe fraction of open base-pairs with varying $T$ at a 
fixed $f=0$ (Figure \ref{contactplot2}). The BD simulations show 
that for homogeneous DNA composed of AT or GC base-pairs, a transition 
occurs from a bound state, where most base-pairs remain intact, to an 
unbound state, where all base-pairs are fully separated. Due to the 
higher base-pair interaction energy of GC pairs, this transition occurs at a larger 
temperature compared to homogeneous DNA made up of AT base-pairs 
only, which has weaker base-pair interaction energy
 (Figure \ref{contactplot2}a). A similar observation is seen 
 from the GNM for homogeneous DNA sequences 
 (Figure \ref{contactplot2}b). 
 For heterogeneous sequences, such as 2G32A30G and 24A32G8A, an intermediate 
 state emerges as the temperature 
$T$ increases, is also in agreement with GNM predictions 
(Figures \ref{contactplot2}a and \ref{contactplot2}b). Notably, 
the presence of AT base-pairs in DNA sequences results in the 
spontaneous opening of a few base-pairs even in 
the absence of an applied force ($f=0$) at 
low temperatures 
in heterogeneous DNA compared to homogeneous 
DNA of GC base-pairs, where base-pair opening is much less prevalent under 
similar conditions (Figure \ref{contactplot2}a and \ref{contactplot2}b). 
The melting temperature of DNA for different sequences can be determined by
monitoring the temperature at which half of the base-pairs become unbound 
at a given 
force \cite{theodorakopoulos2019statistical}. To investigate the synergistic effect of force 
and temperature on DNA melting, we study 
the $f-T$ phase diagram for different DNA 
sequences using BD simulations and the GNM model 
over a range of applied forces $f$ and temperatures $T$.

\subsection{Force-Temperature (\texorpdfstring{$f-T$}{f-T}) diagram for 
DNA sequences with a moving GC block}

\begin{figure}[t]
  \centering 
  {\includegraphics[width=1.1\linewidth,scale=1.8,trim={0.15in 3.8in 0.0in 0.0in},clip]{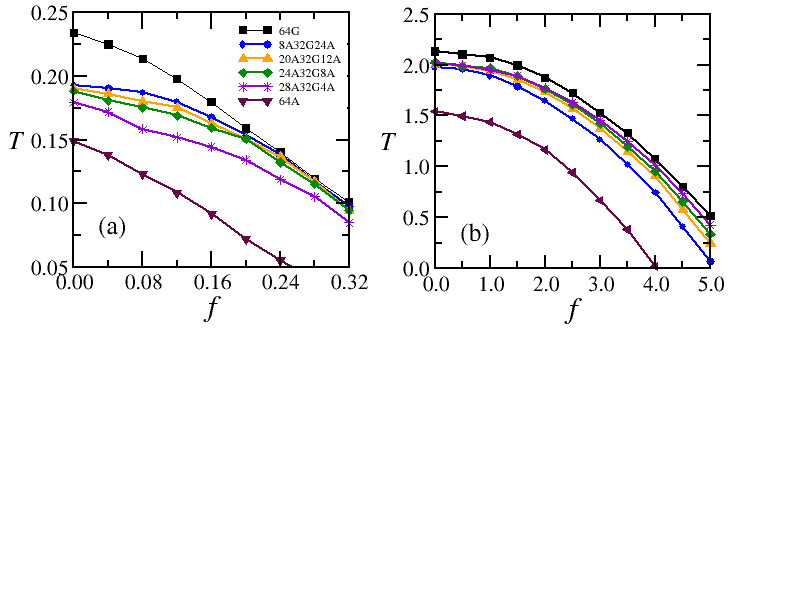}}
  \caption{Force-Temperature ($f-T$) diagram for various sequences 
  with moving GC block from force end to loop end using 
  (a) BD simulation, (b) GNM.}
  \label{FTmovingGC}
\end{figure}

For homogeneous sequences, we observe that the $T_m$ decreases 
with increasing $f$ (Figure \ref{FTmovingGC}, Figure S1). 
Complete melting of DNA composed solely of AT base-pairs 
occurs at relatively low temperatures, while the highest 
melting temperature is observed for DNA consisting 
exclusively of GC base-pairs at different forces. 
This trend is consistently observed across both models
(Figures \ref{FTmovingGC}a and \ref{FTmovingGC}b).
In the heterogeneous case of $50\%$ GC and $50\%$ AT content, 
we first focus on the \texorpdfstring{$f-T$}{f-T} diagram for DNA 
sequences where the position of a 32G block is shifted from the 
force end to the opposite end 
(Figure \ref{bubbbleplot}a). These sequences allow the AT-rich regions 
to form free strands at the force end and the bubble at the opposite end. 
For heterogeneous sequences, the melting temperature falls between that 
of homogeneous AT/GC base-pair of DNA (Figure \ref{FTmovingGC}a 
and \ref{FTmovingGC}b). At $f=0$, the $T_ 
m$ decreases as the 32G block moves farther from the force application end 
(Figure \ref{FTmovingGC}a). However, this observation contradicts 
the predictions of the GNM (Figure \ref{FTmovingGC}b), which suggests 
that a higher $T_m$ is required when the 32GC block is positioned 
further from force end at $f=0$ (blue line with circle and violet 
line with star symbol are in opposite order in Figure \ref{FTmovingGC}a 
and \ref{FTmovingGC}b). To elucidate the origin of this difference, we note that as 
the 32G block shifts farther from the force end, the bubble 
size at the loop end decreases, while the free strand length 
at the force end increases. The presence of longer free strands 
enhances entropy, leading to a reduction in the melting 
temperature ($T_m$), as observed in BD simulations. 
To verify this effect, we examined how the presence of free 
strands at the force end influences melting by adding an 
extra free strand at that force end. BD simulations show 
a reduction in $T_m$ under these conditions, whereas GNM 
displays only a minimal response (Figure S2), suggesting 
that it does not adequately account for the entropic 
contribution of free ends.
\begin{figure}[t]
 \centering 
\includegraphics[width=1.1\linewidth,scale=1.2,trim={0.04in 0.7in 0.0in 1.3in},clip]{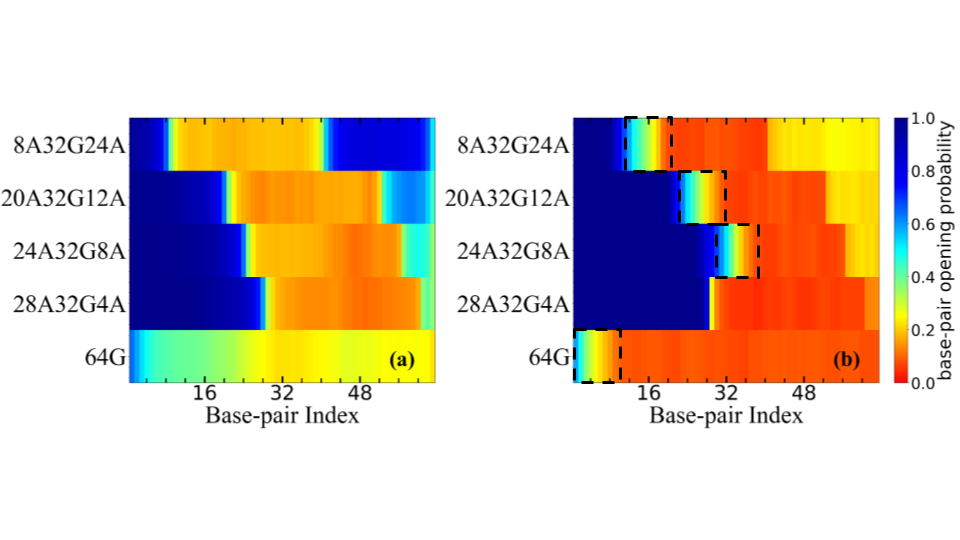}
\caption{Color map showing the opening probability of base-pair 
as a function of base-pair index (x-axis) in different sequences 
(y-axis) just below $T_m$ at (a) $f=0.04$ and (b) $f=0.24$ 
using BD simulation.}
\vspace{-0.15 in}
 \label{probofopenwithseed}
\end{figure}

Another significant difference we observe in BD simulations, 
compared to the GNM, is that at higher forces, for 
heterogeneous sequences where the GC block is positioned 
closer to the force end, the melting temperature ($T_m$) 
is the same as of homogeneous DNA composed entirely of 
GC base-pairs. To gain deeper insight into this behavior, 
we analyzed the microscopic details of base-pair opening. 
At low forces near the melting temperature, we observe 
large base-pair opening probability (indicated by the blue color) 
both at the force end and the opposite end. This 
suggests that base-pairs from both the loop end and the 
force end contribute to DNA opening (Figure \ref{probofopenwithseed}a, 
Figure S3). However, as the applied force increases, the melting 
behavior changes significantly — only the base-pairs at the force 
end participate in the opening (blue color region), while 
the base-pairs at the loop end remain largely intact 
(yellow color at end), unlike the low-force scenario 
(Figure \ref{probofopenwithseed}b, Figure S4). We further 
observe that at high forces near the melting temperature 
(Figure \ref{probofopenwithseed}b), for homogeneous DNA of 
GC base-pairs, melting is initiated by the stretching of only a 
few GC base-pairs ($\approx 9$) near the force end, after 
which a slight 
increase in temperature leads to the separation of all base-pairs. 
Similarly, for all heterogeneous sequences where a GC block 
follows a few AT base-pairs near the force end, the AT 
base-pairs open at relatively lower temperatures, while the
GC block acts as a barrier. Initiating the melting of the DNA 
chain requires the stretching of $\approx 9$ GC base-pairs 
(indicated by the region in black rectangular box in the 
Figure \ref{probofopenwithseed}b), regardless of the sequence 
that follows after GC. This behavior explains why 
the \texorpdfstring{$f-T$}{f-T} curves of heterogeneous 
sequences merge with those of homogeneous GC sequences at 
high forces. However, in the case of GNM (Figure \ref{FTmovingGC}b), 
there is no prescription to capture stretching of 
base-pairs, 
the base-pairs can be either open or closed. 
Hence no such characteristic length scale is observed in 
GNM (Figure \ref{FTmovingGC}b), as seen in BD simulations (Figure \ref{FTmovingGC}a). However, the emergence of a characteristic 
length from BD simulations align with earlier observations that 
suggest the existence of a threshold length in DNA unzipping 
beyond which the applied force no longer influences the base 
pairs \cite{singh2016opening}.

\subsection{Force-Temperature (\texorpdfstring{$f-T$}{f-T}) 
diagram for DNA sequences with a moving AT block}  
We further examine the (\texorpdfstring{$f-T$}{f-T}) diagram 
for DNA sequences where a 32A block is shifting from the force 
end to the opposite end. In these sequences, the AT-rich regions 
form a loop constrained between strongly interacting GC base-pairs, 
with the loop position varying along the DNA chain for different sequences. 
Figure \ref{FTmovingAT} (and Figure S5) shows the variation of $T$ as 
a function of  $f$.

Interestingly, no inversion is observed between the BD simulation 
and GNM results in the case of a moving 32A block (blue line with 
circle and purple line with star symbol in Figures \ref{FTmovingAT}a 
and \ref{FTmovingAT}b respectively), unlike the behavior seen when a 
32G block was shifted along the DNA chain (blue line with circle and 
purple line with star symbol in Figures \ref{FTmovingGC}a and 
\ref{FTmovingGC}b respectively). For the moving 32A block, where 
the loop is confined between strongly interacting GC regions, 
the constraint in the GNM associated with accurately accounting 
for free-strand entropy becomes unimportant, as there is no free 
strands at the force end. The GC base-pair stabilizes the end. 
In both models, when the loop is positioned closer to the force end, 
the structure exhibits lower stability compared to configurations 
where the loop is located farther from the force end. At low forces, 
the 32A block adopts a loop configuration, and the associated entropy 
facilitates the opening of the base pair at the loop’s end. For a 
smaller number of GC base-pairs at the force end, melting of the 
GC base-pairs near the force end occurs at a lower temperature, 
as the combined effects of loop entropy and the entropy of the 
stretched free strand promoting destabilization of GC base-pairs 
at the force end. For larger number of GC base-pairs at force end, 
as the melting of DNA requires $50\%$ base-pair opening and a loop
of 32A is already contribute as open base-pairs. It requires only few 
GC base-pair opening at force end. Therefore, temperature required to 
melt 16G32A16G and 28G32A4G are very close. This consistent trend indicates 
that both models capture the effects of loop entropy accurately (Figure S6). 

At high forces, the BD simulations show a consistent behavior regardless 
of whether the moving block is AT or GC, in DNA sequences containing 
$50\%$ AT and $50\%$ GC content. When the AT block is positioned very 
close to the force end, following a 4G or 8G segment, the melting temperature 
still differs from that of homogeneous GC DNA. At a fixed $f$, $T_m$ is 
lower when the AT block follows a 4G segment compared to when it follows an 
8G segment. This difference arises because fewer GC base-pairs require a 
lower melting temperature, which is sufficient to break additional AT 
base-pairs and ultimately leads to the melting of half the DNA chain.
Furthermore, when the AT block is positioned at force end, following a 
GC segment $> 8$ base-pairs, the $f-T$ diagram of the heterogeneous sequence 
closely aligns with that of homogeneous DNA of GC base-pairs 
(Figure \ref{FTmovingAT}a). This result further supports the 
conclusion that, at high forces, the stretching  of $\approx 9$ 
GC base-pairs is sufficient to separate the two DNA strands. 
The \texorpdfstring{$f-T$}{f-T} diagrams show overall agreement 
between the two models as the bubble moves from one end to the other. 
However, GNM lacks a characteristic length scale, which arises from its 
binary representation of hydrogen bonds as either fully open or fully closed.

\vspace{1 cm}
\section*{Conclusion}

\begin{figure}[t!]
 \centering 
  {\includegraphics[width=1.1\linewidth,scale=1.8,trim={0.15in 3.8in 0.0in 0.0in},
  clip]{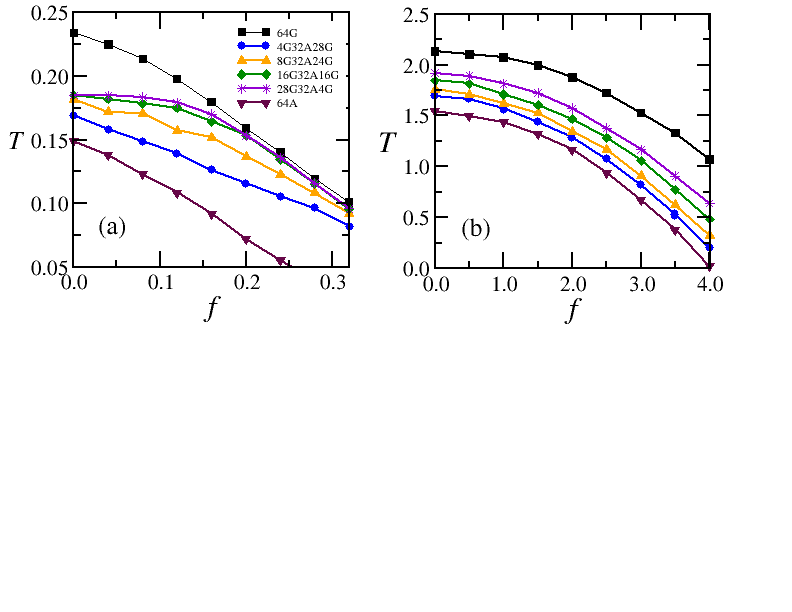}}
  \caption{Melting Force-Temperature profile for different sequences with 
  moving AT block using (a) BD simulation (b) GNM.}
   \label{FTmovingAT}
\end{figure}

 This study examines the role of implicit bubble, arising from base-pair 
 heterogeneity, in the force-induced melting of DNA. To investigate this 
 phenomenon, two distinct modeling frameworks are employed: Brownian 
 Dynamics (BD) simulations, which capture the time-dependent dynamics 
 of DNA unzipping under force, and the Gaussian Network Model (GNM), 
 which relies on normal mode analysis derived from equilibrium 
 fluctuations of DNA nucleotides under force. Despite their 
 methodological differences, both models show consistent qualitative 
 trends-most notably, the observation that the temperature required 
 to separate the DNA strands decreases with increasing applied force. 
 This agreement suggests that key features of DNA melting can be 
 captured even with differing levels of model resolution and complexity.
 
 The major qualitative difference between the results of the two models, 
 lies in the order of stability in the \texorpdfstring{$f-T$}{f-T} 
 curve for sequences with moving GC, (Figure \ref{FTmovingGC}) which 
 is opposite in nature in the two models - BD and GNM. This is because 
 of a lack of proper incorporation of free strand entropy (that might 
 be at play when the DNA is partially denatured at force end) in GNM. 
 BD simulations show a length scale \cite{singh2023pulling} at high 
 $f$ and low $T$ (Figure \ref{FTmovingGC}a)which signifies that 
 breaking of $\approx 9$ of GC base pairs would break the rest of 
 the sequence irrespective of the information of heterogeneity in 
 the rest of the sequence. The presence of such a length scale is 
 also there in the BD simulation results of \texorpdfstring{$f-T$}{f-T} 
 curves for sequences with moving AT block(Figure \ref{FTmovingAT}a). 
 However, in the case of sequences with a moving AT block 
 (Figure \ref{FTmovingAT}), the force–temperature (\texorpdfstring{$f-T$}{f-T}) 
 phase diagrams obtained from both models show good qualitative agreement, 
 suggesting that bubble formation is effectively captured in both approaches.  

Our study also highlights the influence of base stacking interaction 
on the force-temperature (\texorpdfstring{$f-T$}{f-T}) phase diagram of 
DNA. It is also important to note that the GNM model is optimized by 
incorporating additional energy terms, such as stacking interactions 
between DNA bases derived from quantum calculations. However, these 
stacking interactions added beyond the base-pairing terms do not lead 
to any significant qualitative difference when compared to the BD 
simulations, which exclude them. This suggests that stacking effects 
may be effectively captured through base-pairing interactions alone.

\section*{Acknowledgements}
We acknowledge the High-Performance Computing Cluster at Ashoka. BM gratefully 
acknowledges the financial support of Ashoka University and DST/INSPIRE (IF210718). 

\bibliographystyle{apsrev4-2}
\bibliography{rsc}

\widetext
\pagebreak
\begin{center}

  \textbf{\large Supplementary Material\\
  }

  \vspace{0.3cm}

\begin{center}
{\large{\bf  ``Role of bubble positioning in force induced melting of DNA''}}\\
{Bidisha Mukherjee$^1$}, {Amit Raj Singh$^2$} and {Garima Mishra$^1$}\\
{$^1$Department of Physics,  Ashoka University, Sonepat 131029, India\\$^2$Graphic Era Hill University, Dehradun, India \\}

\end{center}

  \vspace{0.2cm}

\end{center}

\setcounter{equation}{0}
\setcounter{figure}{0}
\setcounter{table}{0}
\setcounter{page}{1}
\setcounter{section}{0}

\renewcommand{\theequation}{S\arabic{equation}}
\renewcommand{\thefigure}{S\arabic{figure}}
\renewcommand{\thesection}{S\arabic{section}}

\newpage

\begin{figure}[t]
  \centering
  \includegraphics[width=.49\linewidth]{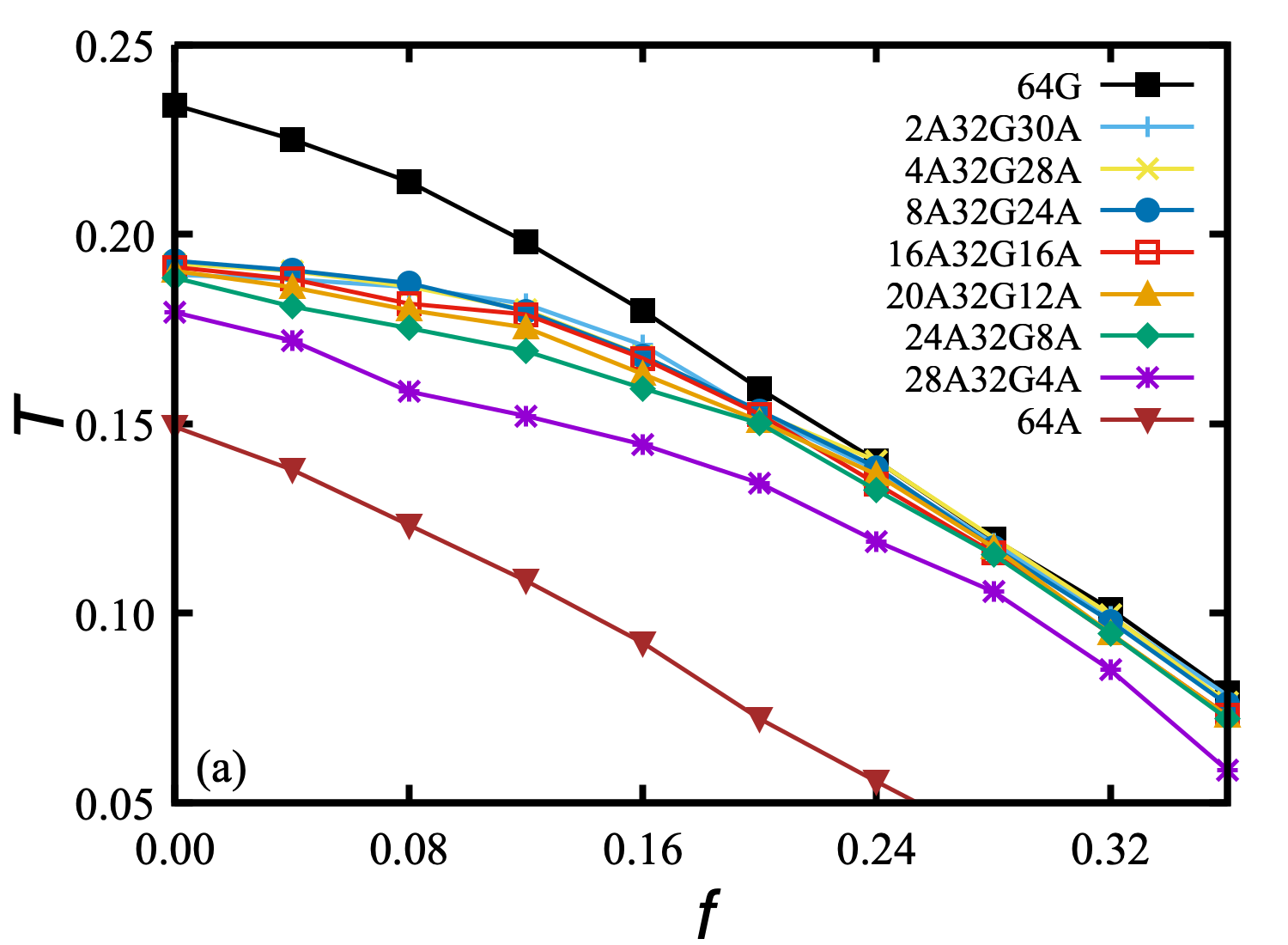}
  \includegraphics[width=.49\linewidth]{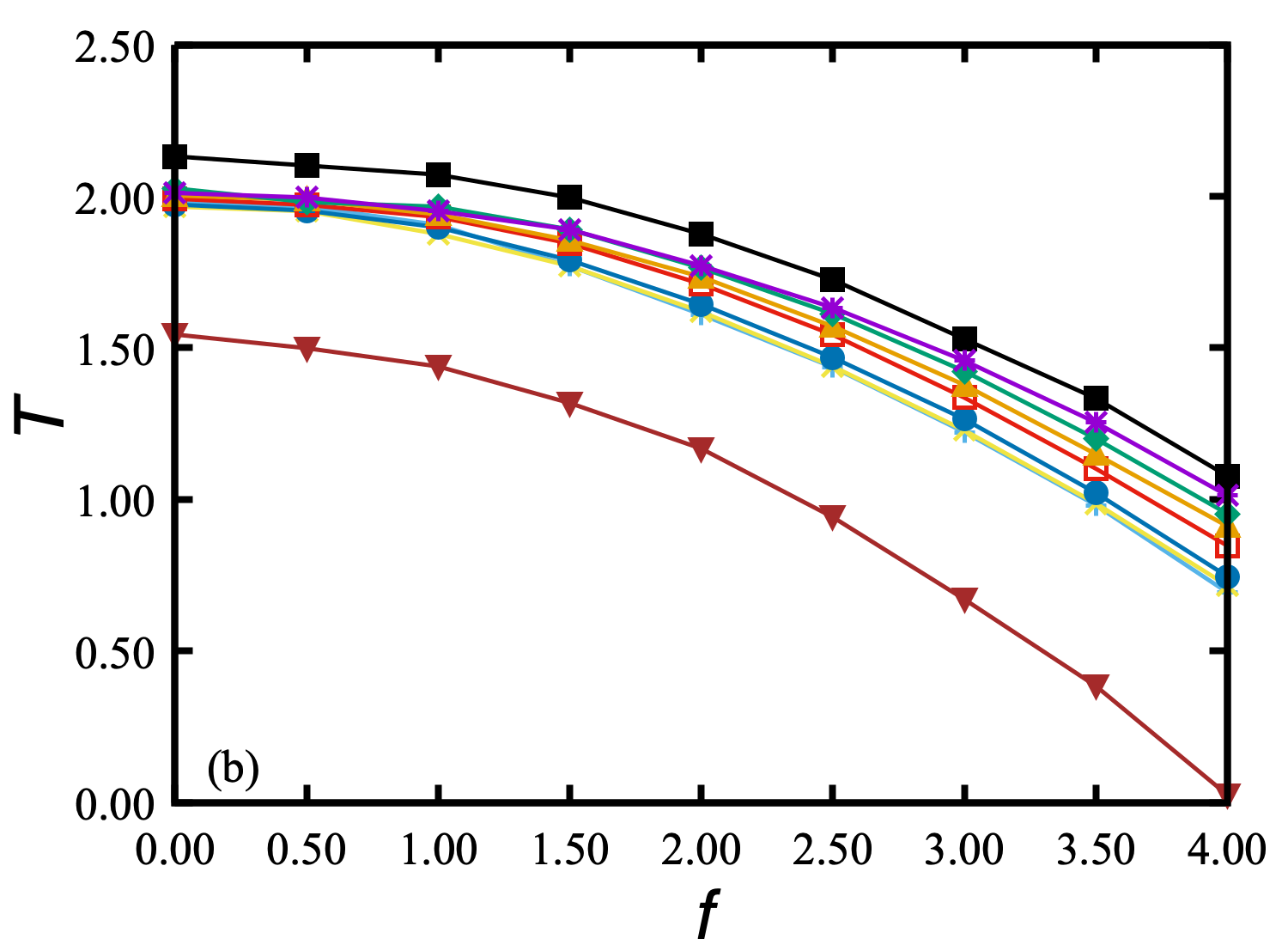}
  \caption{$f-T$ for different sequences with moving GC block from the force end to the other end using (a) BD simulation (b) GNM.}
  \label{fig:AllAs}
\end{figure}

\begin{figure}[t]
 \centering 
  {\includegraphics[width=.49\linewidth]{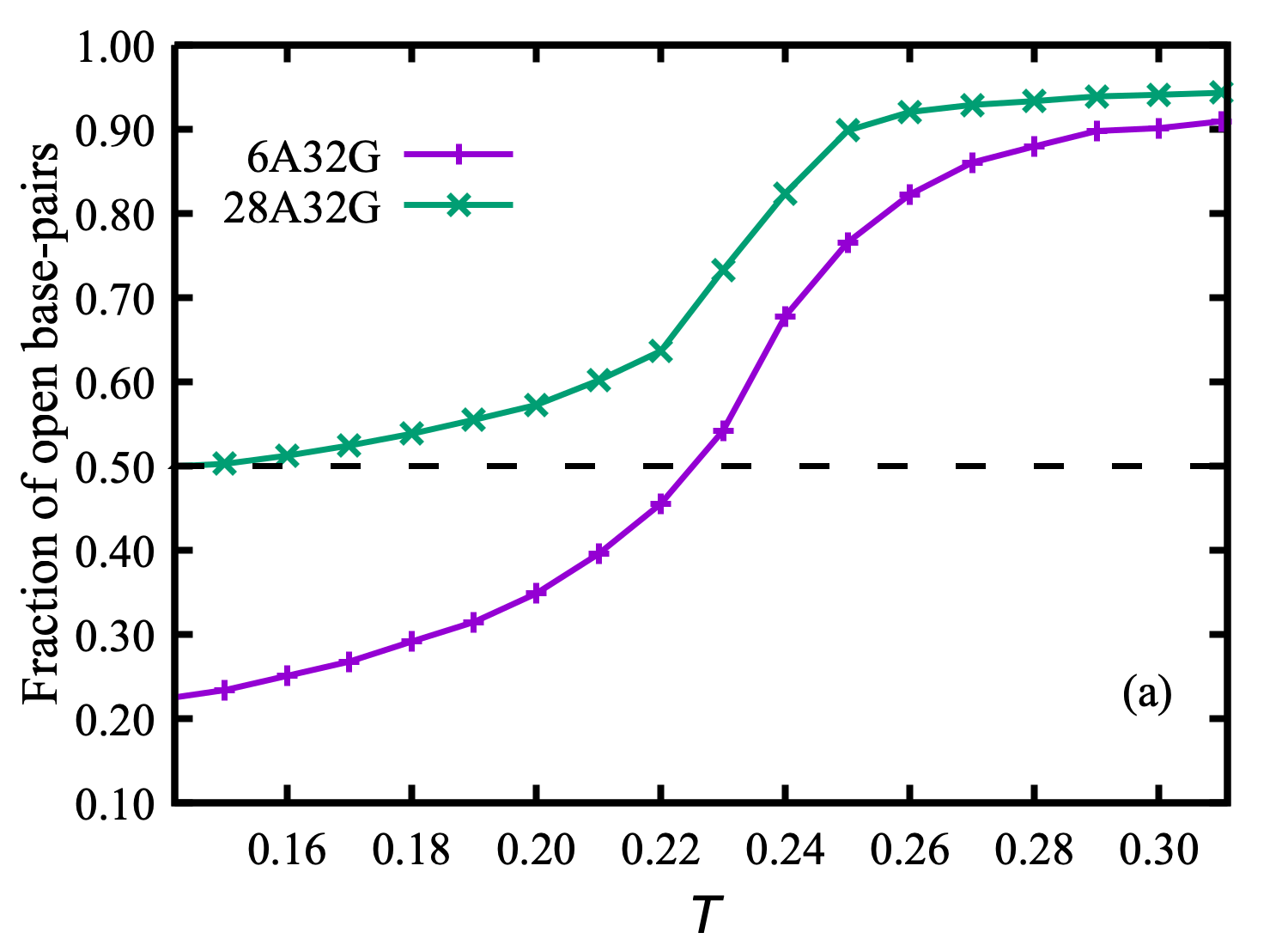}}
   {\includegraphics[width=.49\linewidth]{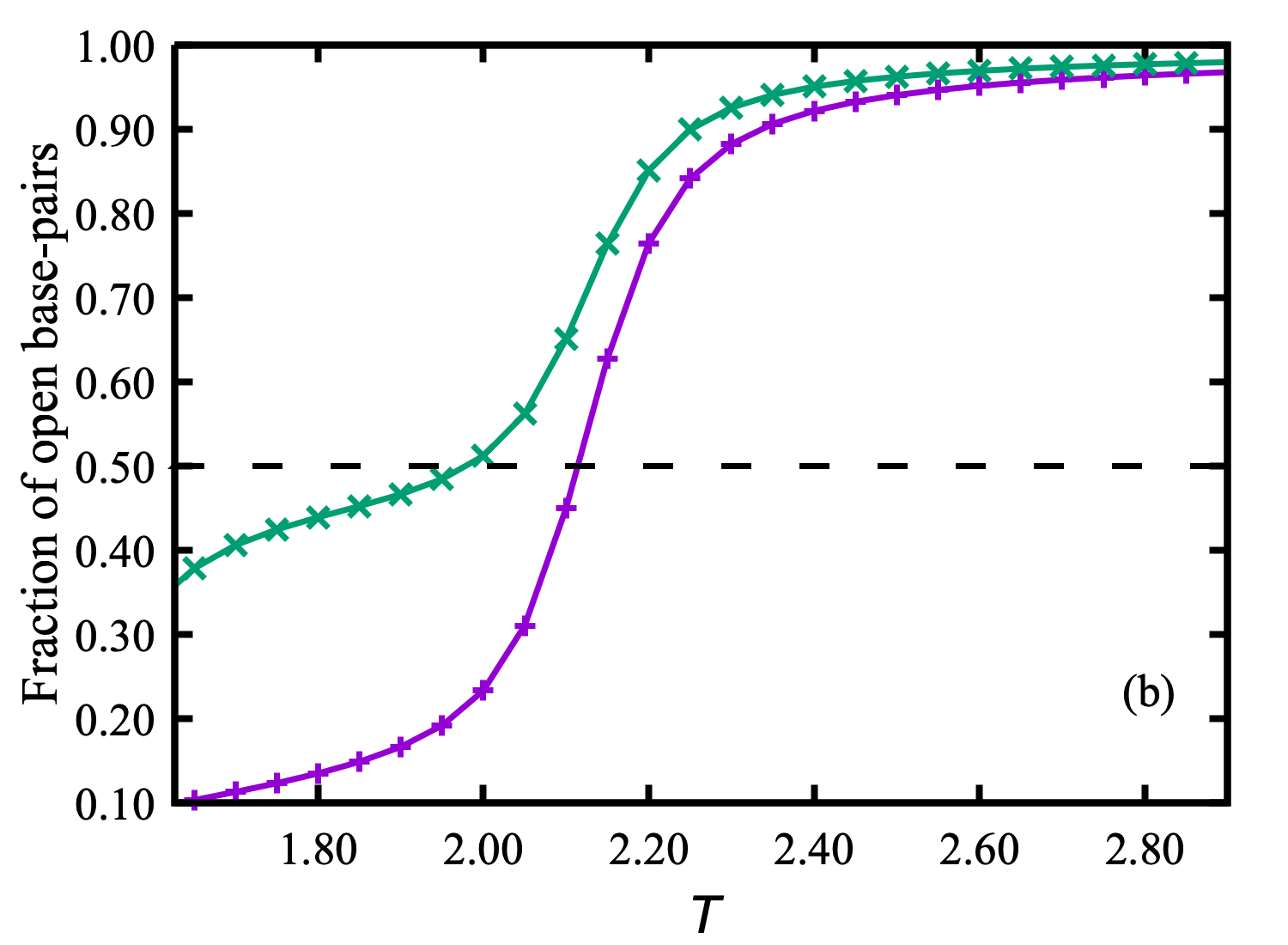}}

  \caption{ Variation of fraction of open base-pairs with temperature at $f=0$ using (a) BD simulation, and (b) GNM. It shows that the effect of free strands is not similar in the two models. The bond strength of AT base-pairs was set to zero to ensure identical enthalpy for DNA chains of varying lengths and to isolate the effect of free strands only.}
  \label{ single strand entropy }
\end{figure}

\begin{figure}[t]
 \centering 
  {\includegraphics[width=.4\linewidth]{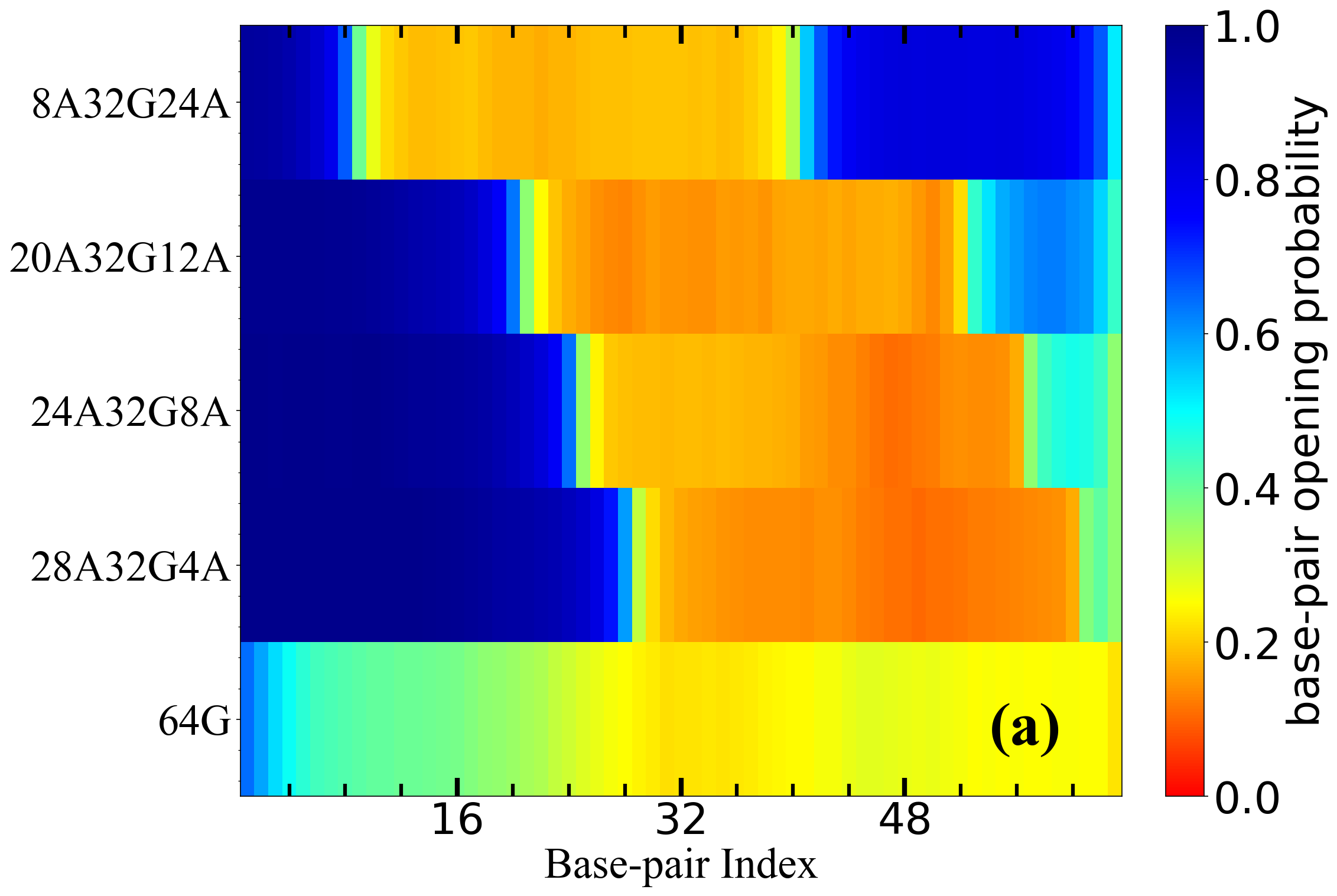}}
  {\includegraphics[width=.4\linewidth]{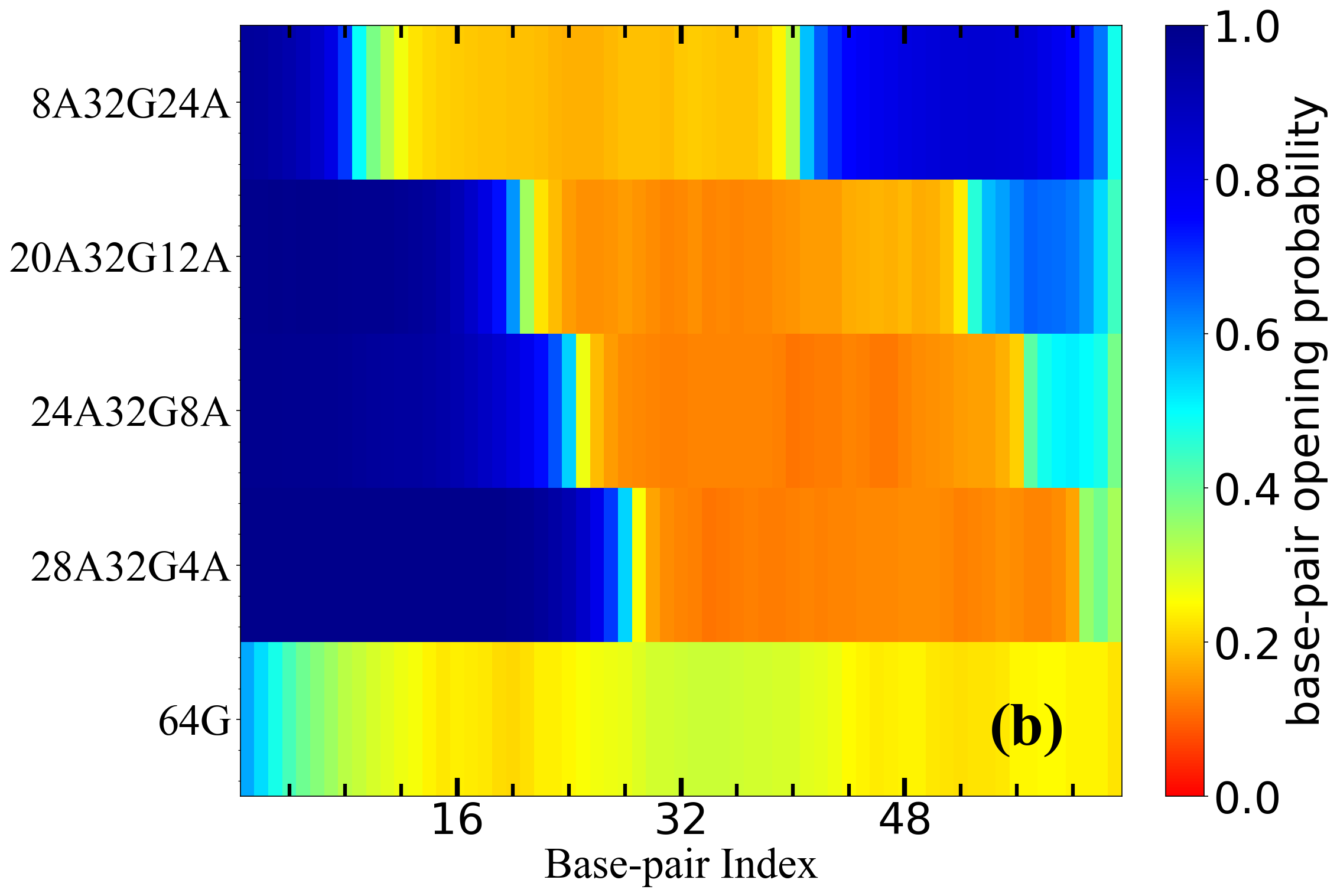}}
  {\includegraphics[width=.4\linewidth]{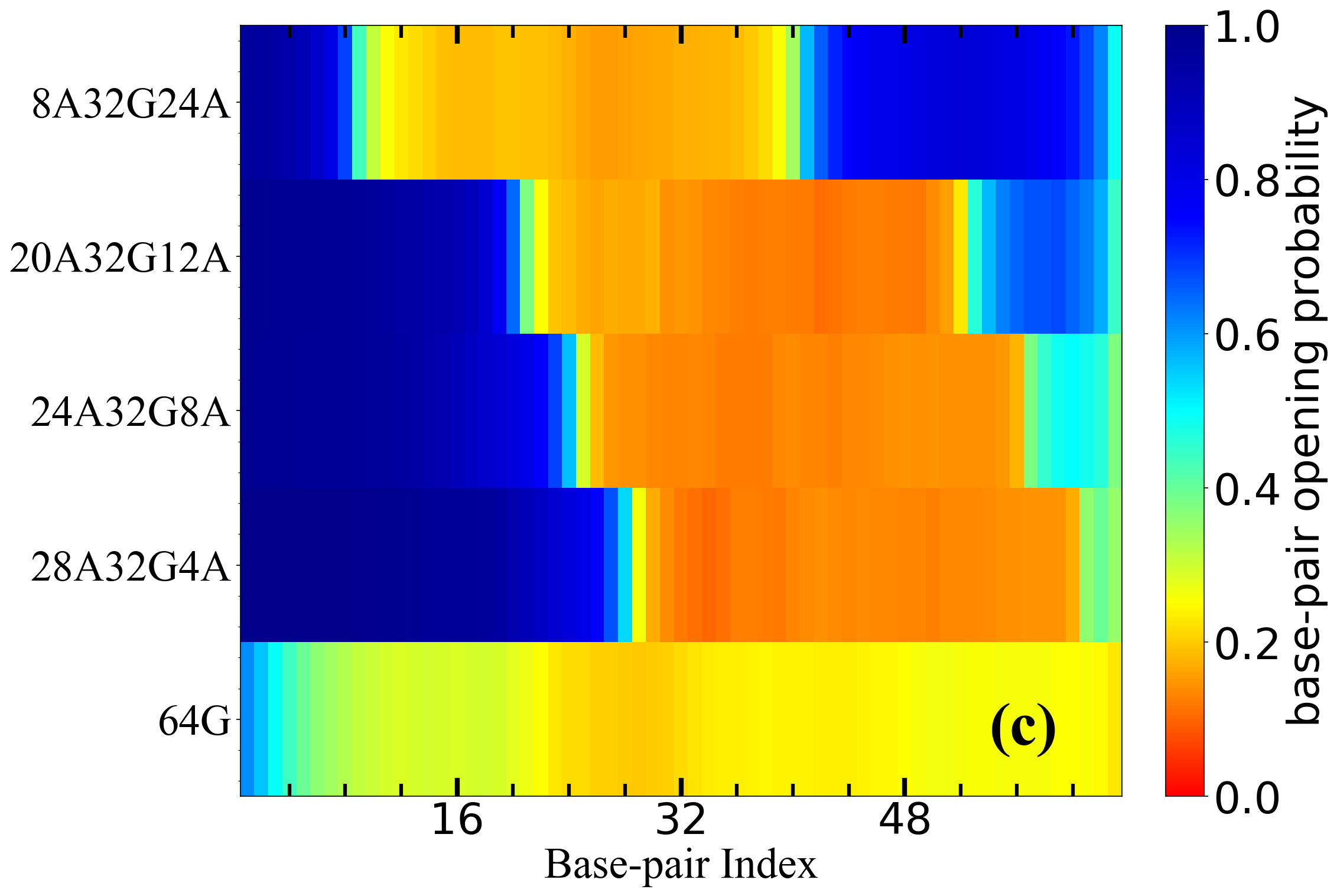}}
  {\includegraphics[width=.4\linewidth]{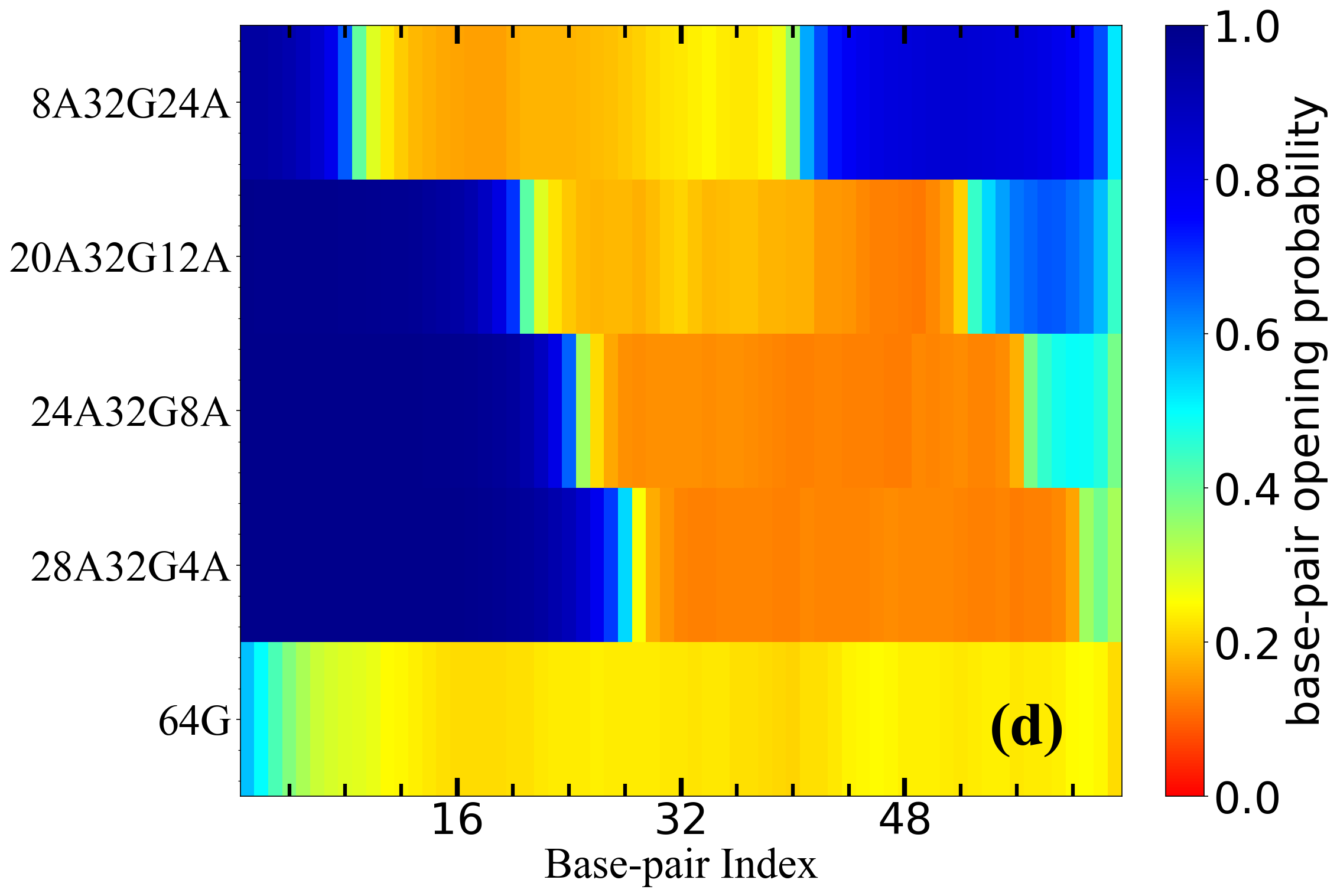}}
  {\includegraphics[width=.4\linewidth]{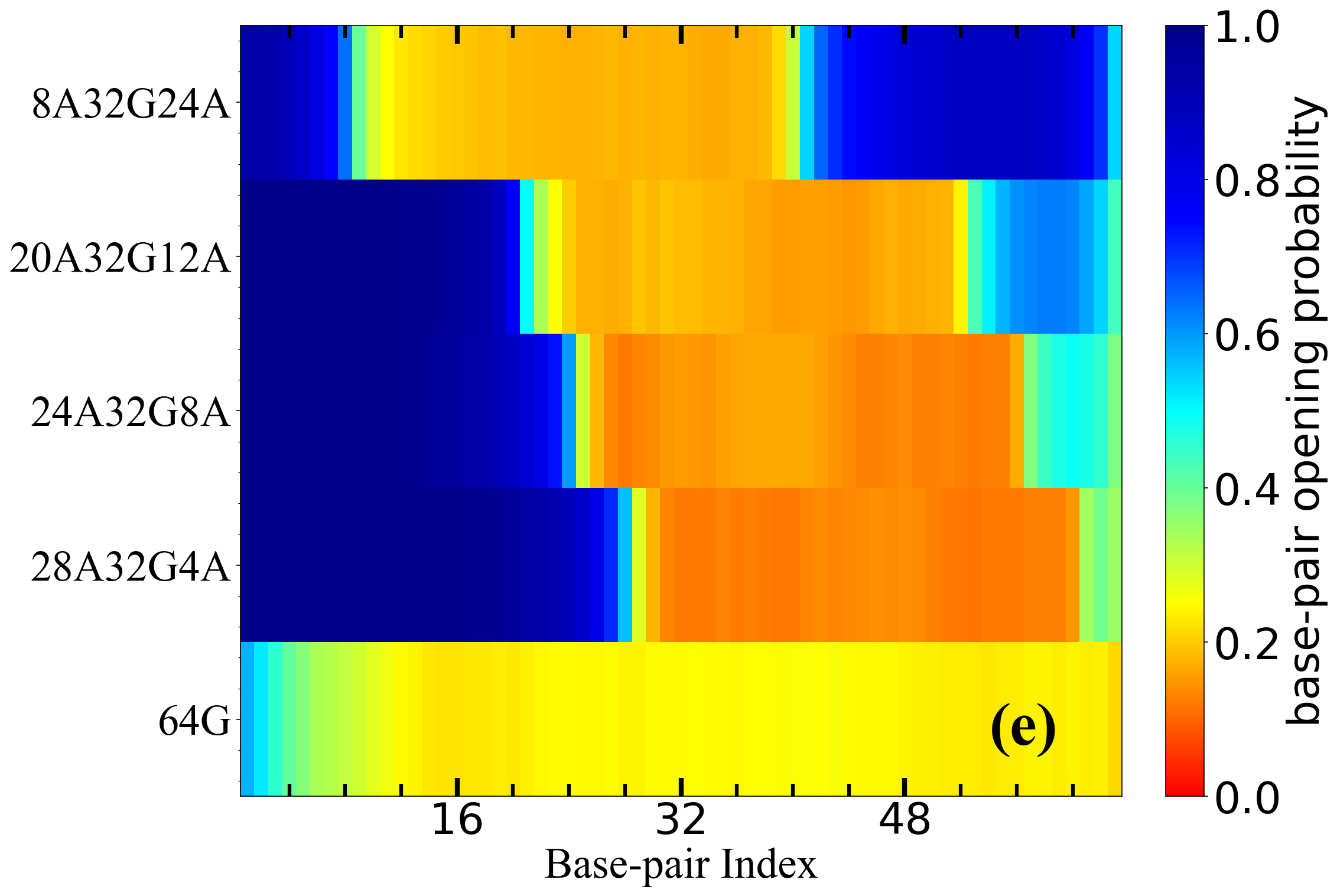}}
  {\includegraphics[width=.4\linewidth]{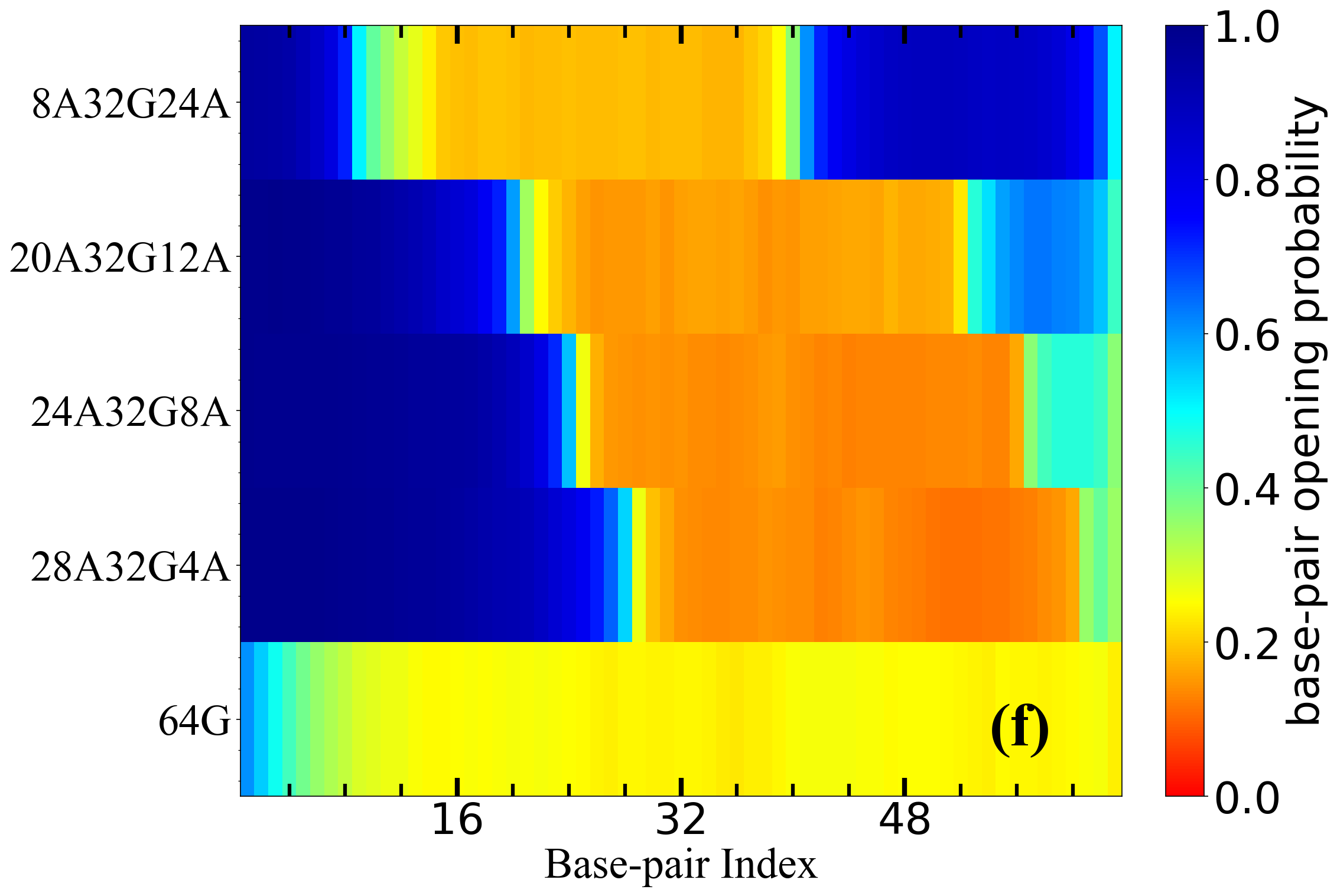}}
  \caption{Probability of base-pair opening at force $f=0.04$, slightly below the respective melting temperatures. At $f=0.04$, the melting temperatures $T_m$ for the sequences  8A32G24A, 20A32G12A, 24A32G8A, 28A32G4A, 64G are 0.190, 0.186, 0.180, 0.171 and 0.225, respectively. Different plots correspond to different random seed values.}
  \label{f0.04}
\end{figure}

\begin{figure}[t]
 \centering 
  {\includegraphics[width=.4\linewidth]{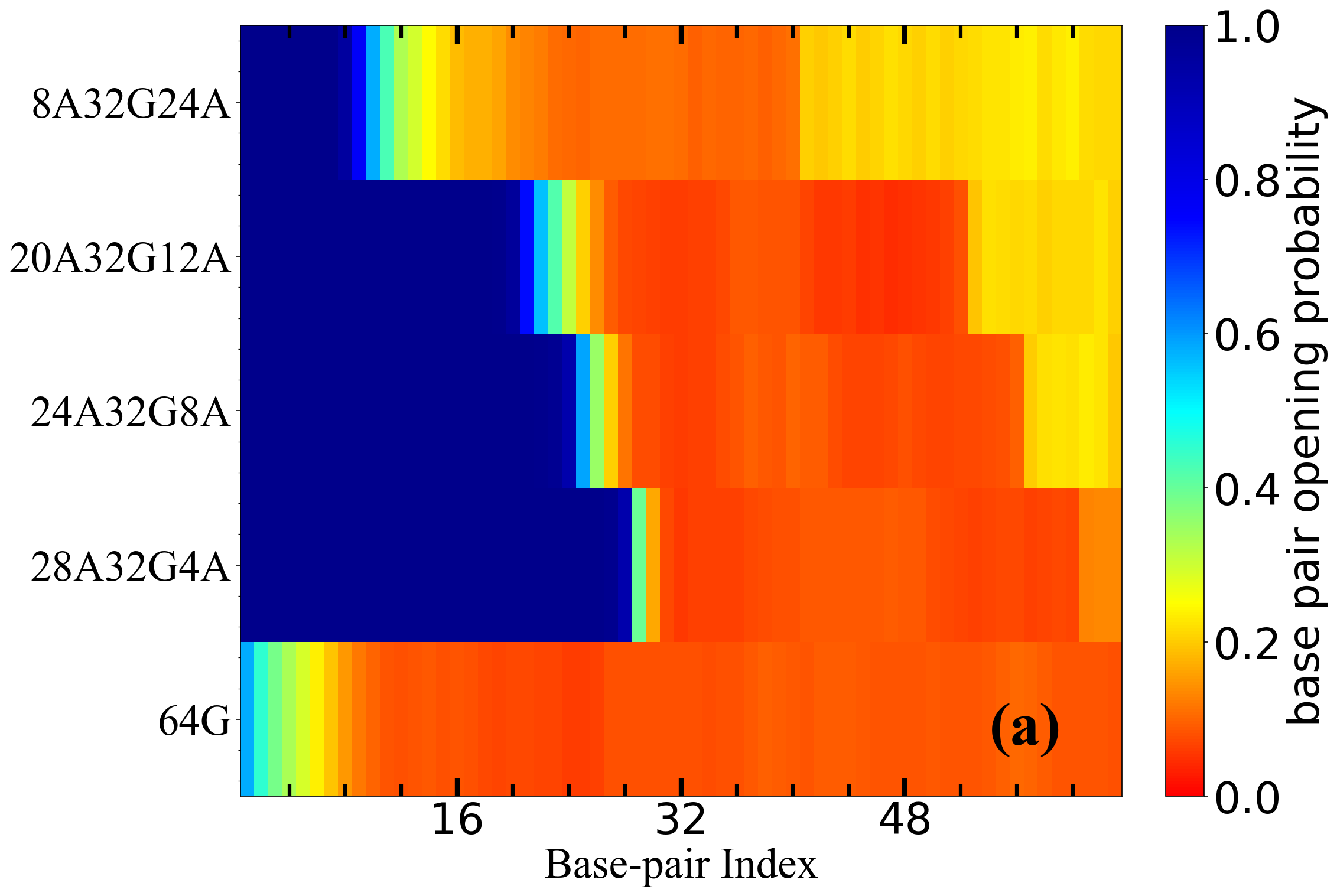}}
  {\includegraphics[width=.4\linewidth]{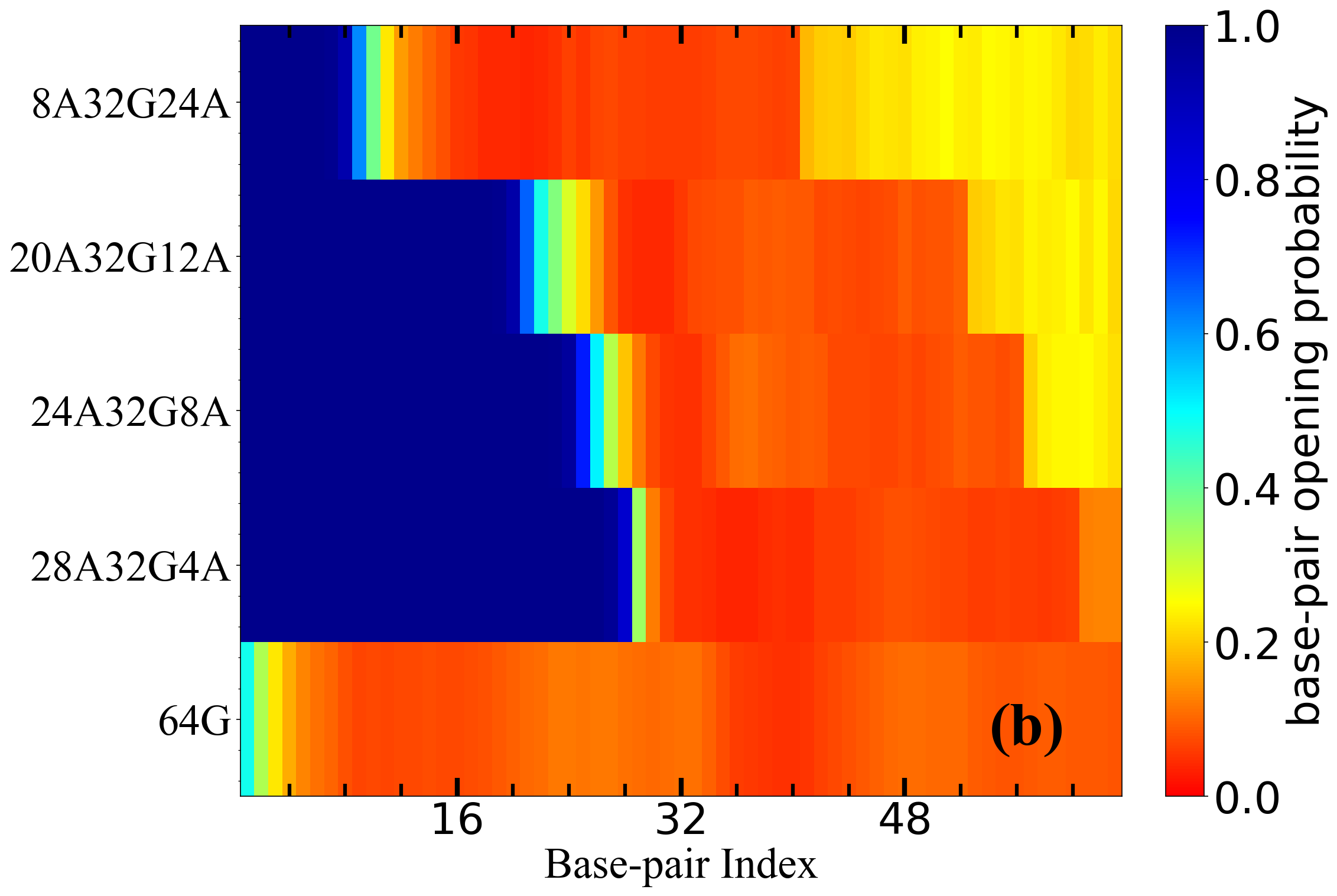}}
  {\includegraphics[width=.4\linewidth]{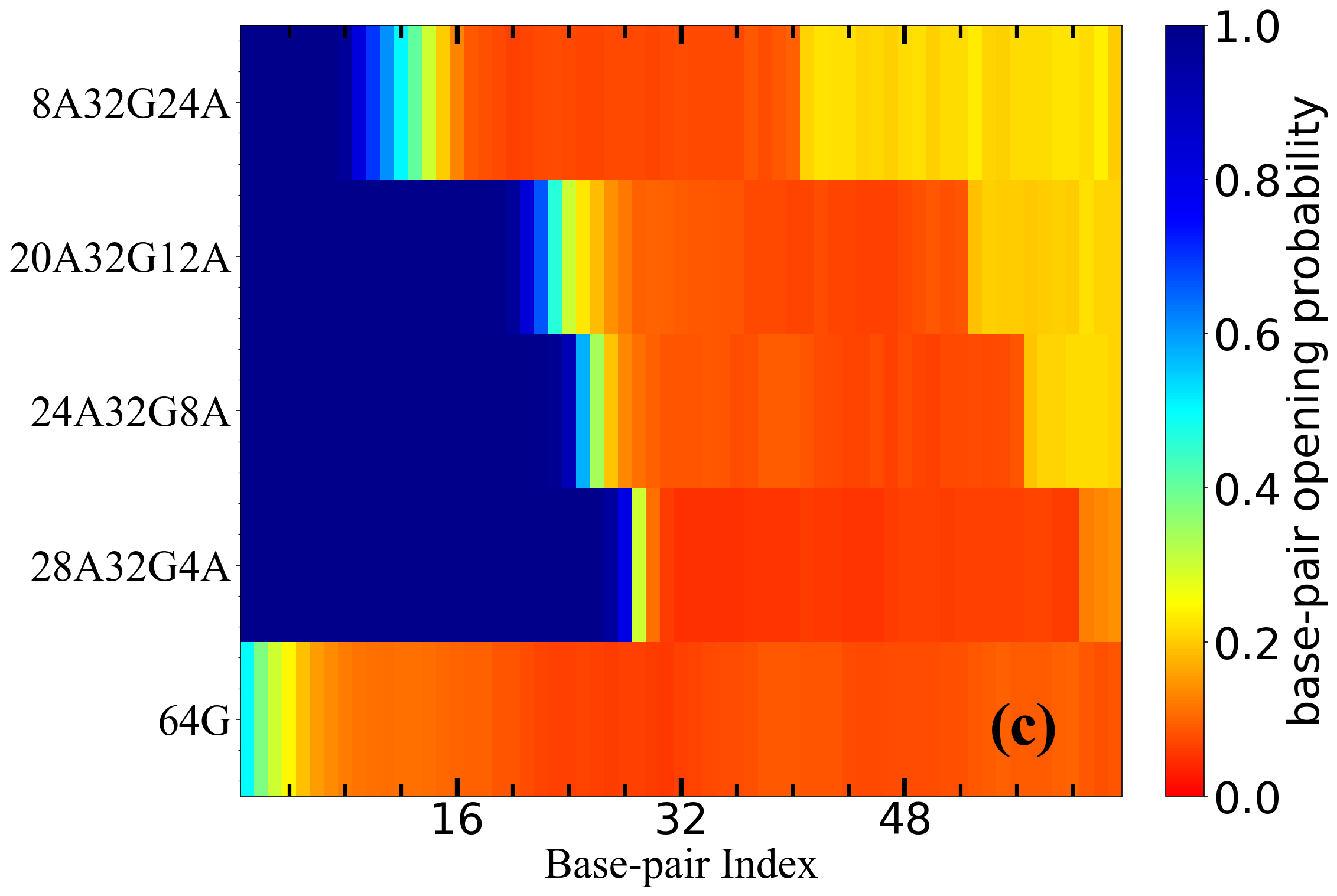}}
  {\includegraphics[width=.4\linewidth]{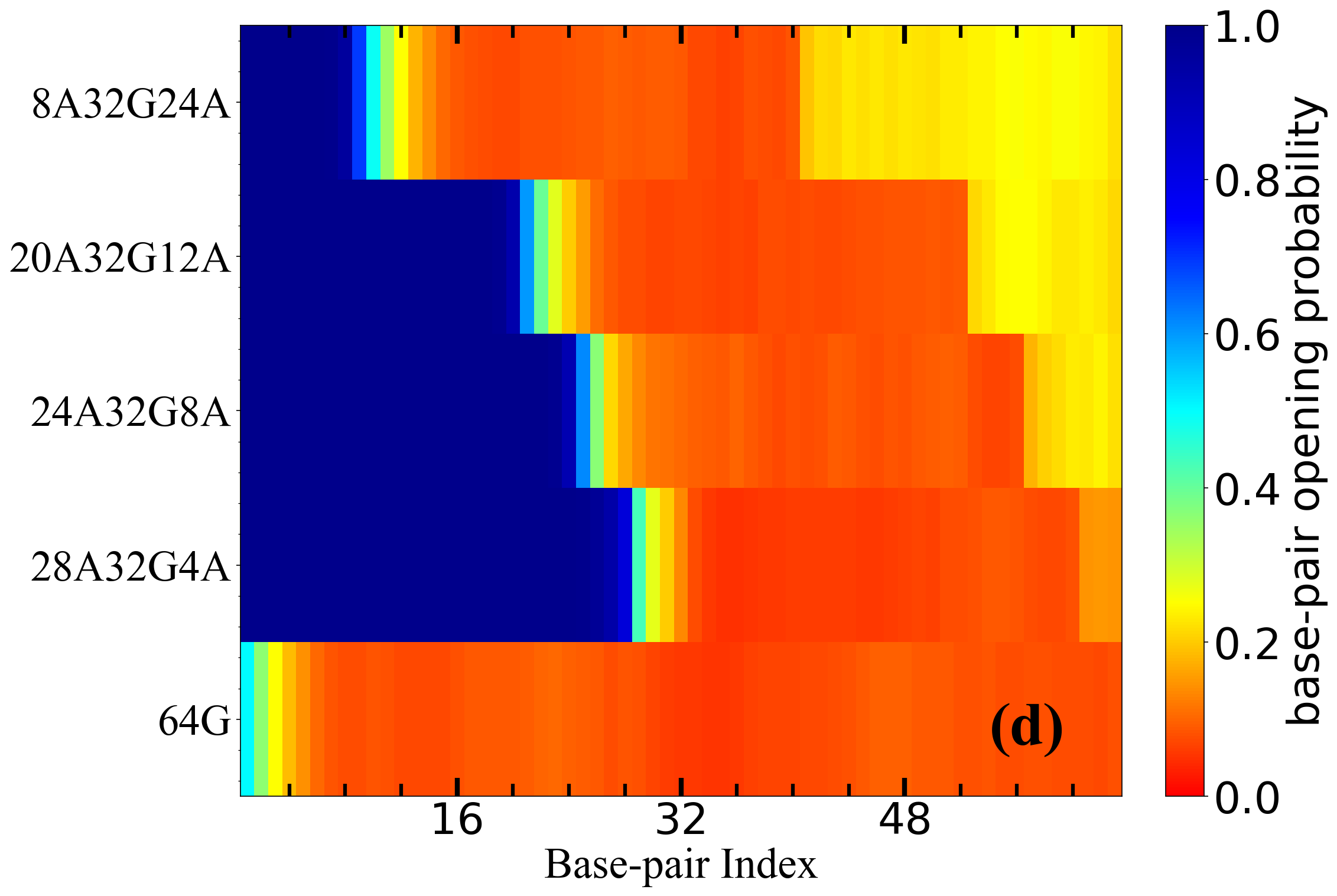}}
  {\includegraphics[width=.4\linewidth]{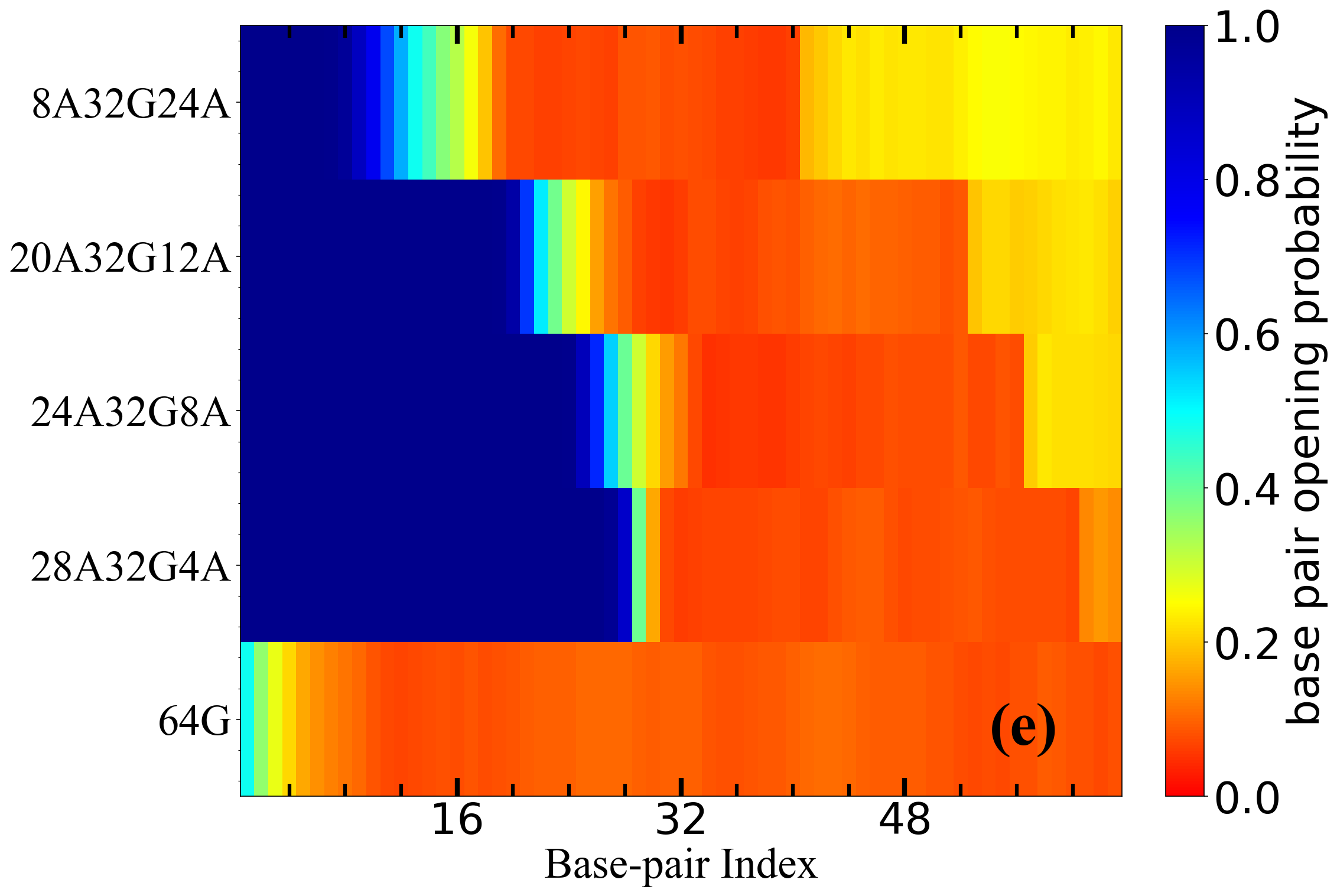}}
  {\includegraphics[width=.4\linewidth]{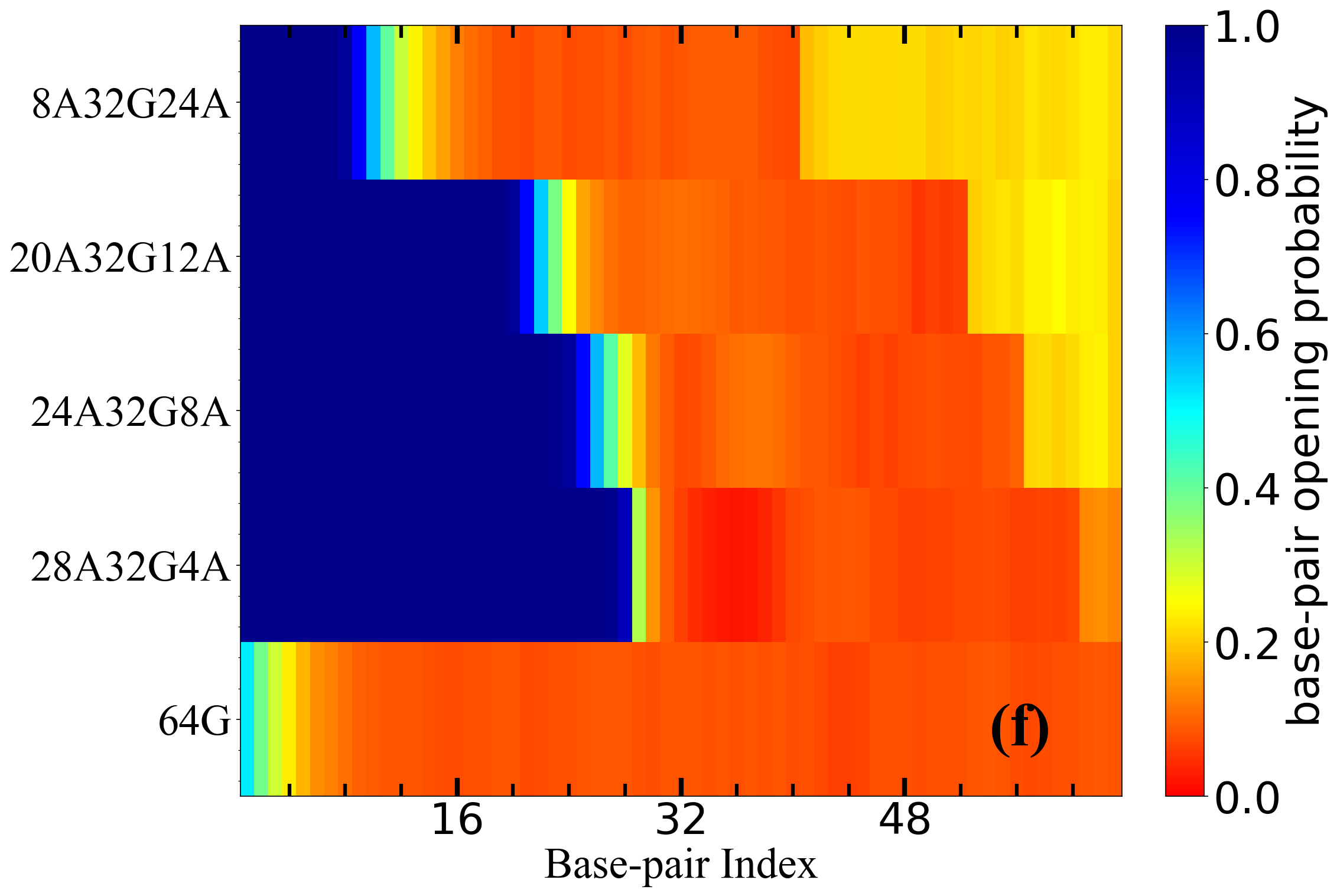}}
  \caption{Probability of base-pair opening at force $f=0.24$, slightly below the respective melting temperatures. At $f=0.24$, the melting temperatures $T_m$ for the sequences 8A32G24A, 20A32G12A, 24A32G8A, 28A32G4A, and 64G are 0.138, 0.136, 0.132, 0.118, and 0.140, respectively. Different plots correspond to different random seed values.}
  \label{f0.24}
\end{figure}

\pagebreak

\begin{figure}[t]
 \centering 
     {\includegraphics[width=.49\linewidth]{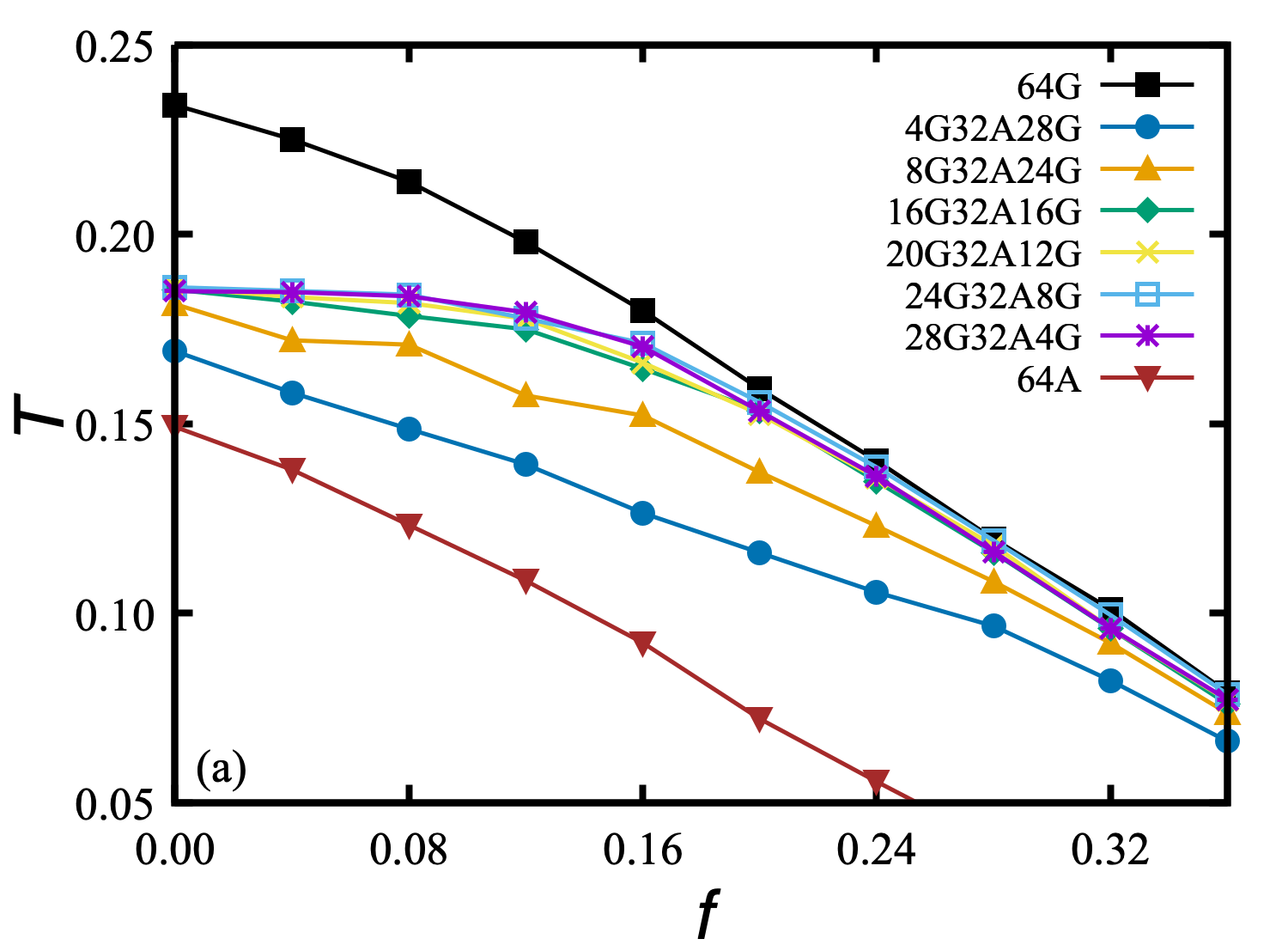}}
    {\includegraphics[width=.49\linewidth]{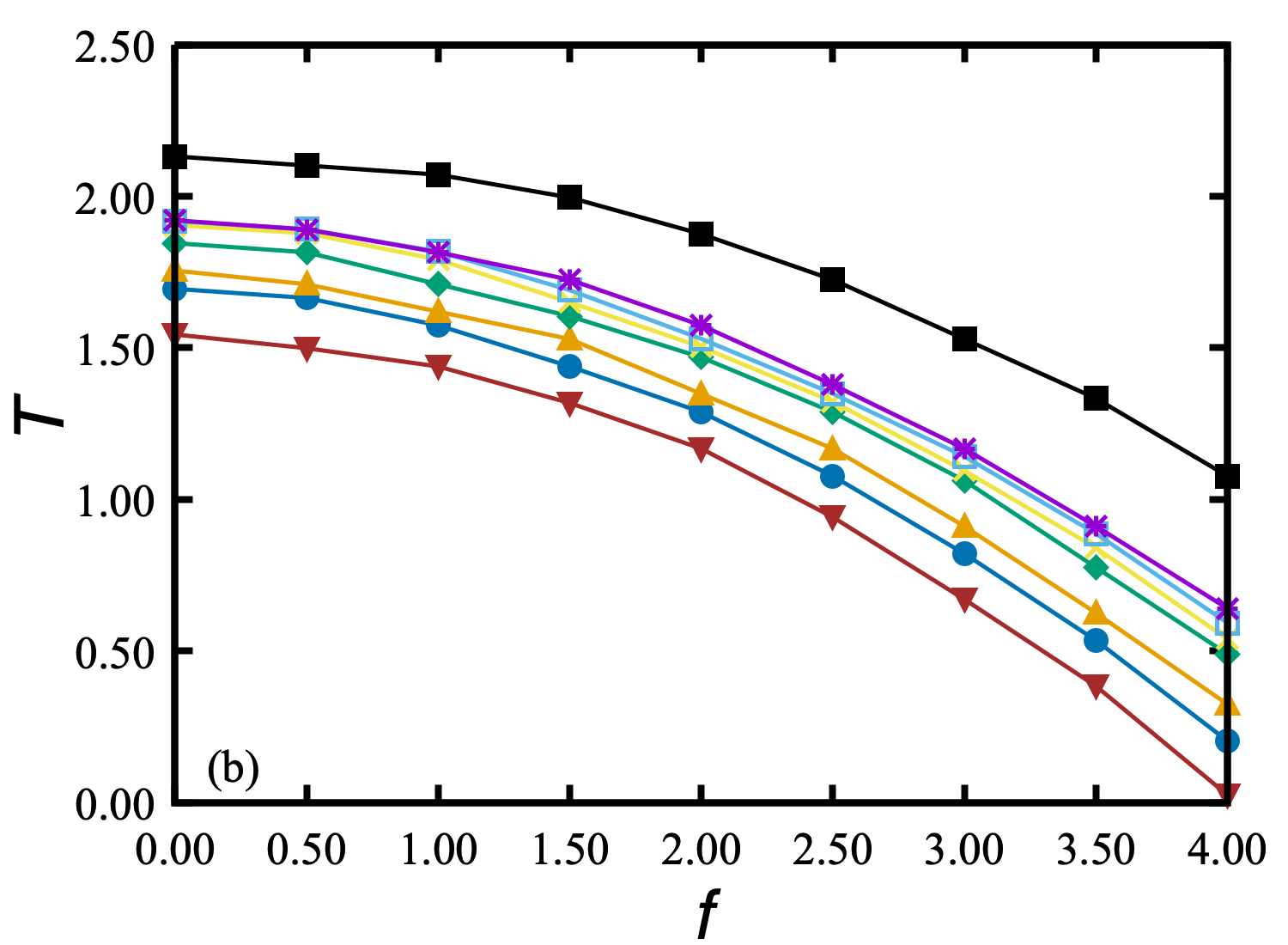}}
  \caption{ \texorpdfstring{$f$-$T$}{f-T} for different sequences with moving AT block from the force end to the other end using (a) BD simulation (b) GNM. }
  \label{ All Gs }
\end{figure}

\begin{figure}[htb]
 \centering 
  {\includegraphics[width=.49\linewidth]{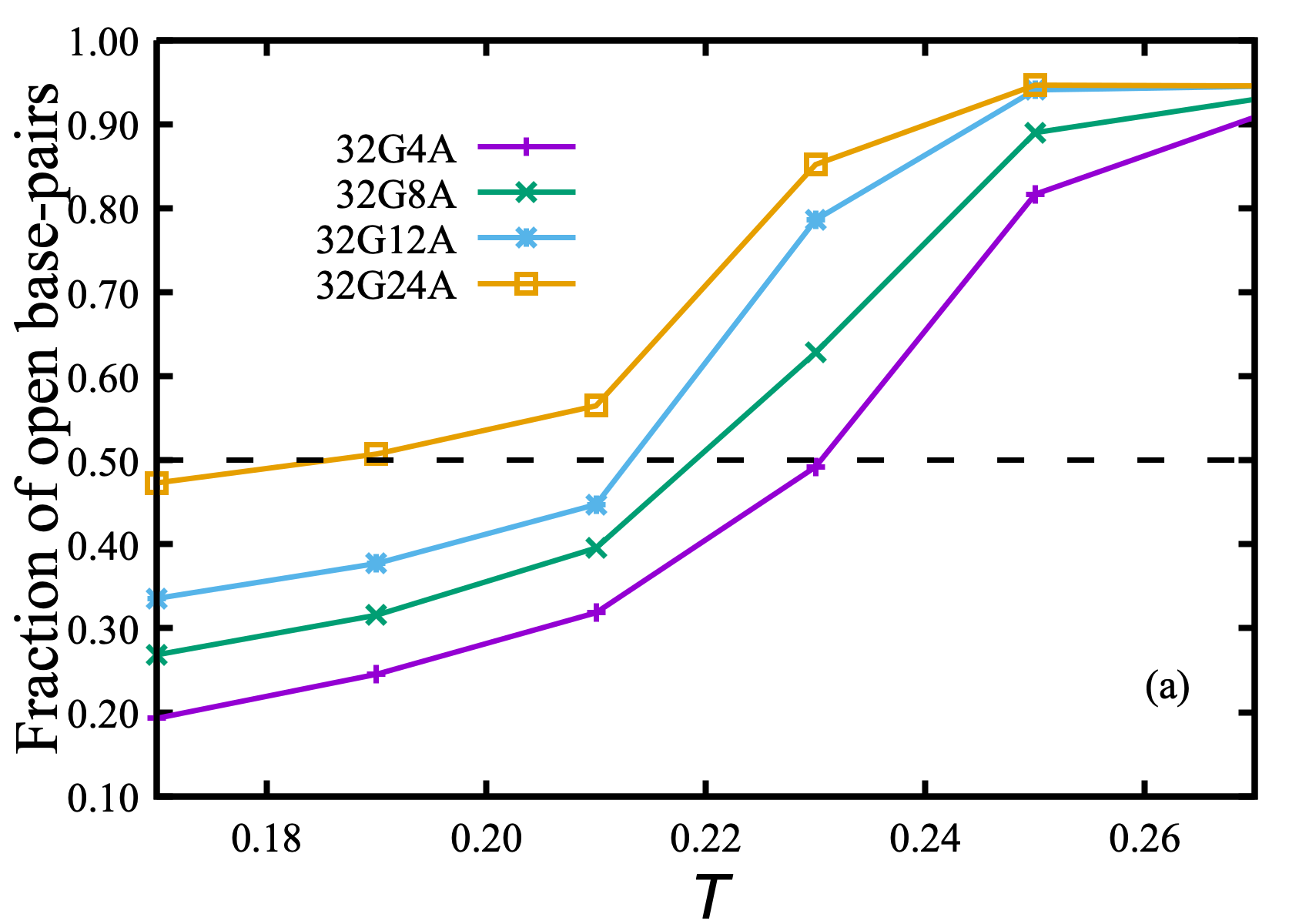}}
  {\includegraphics[width=.49\linewidth]{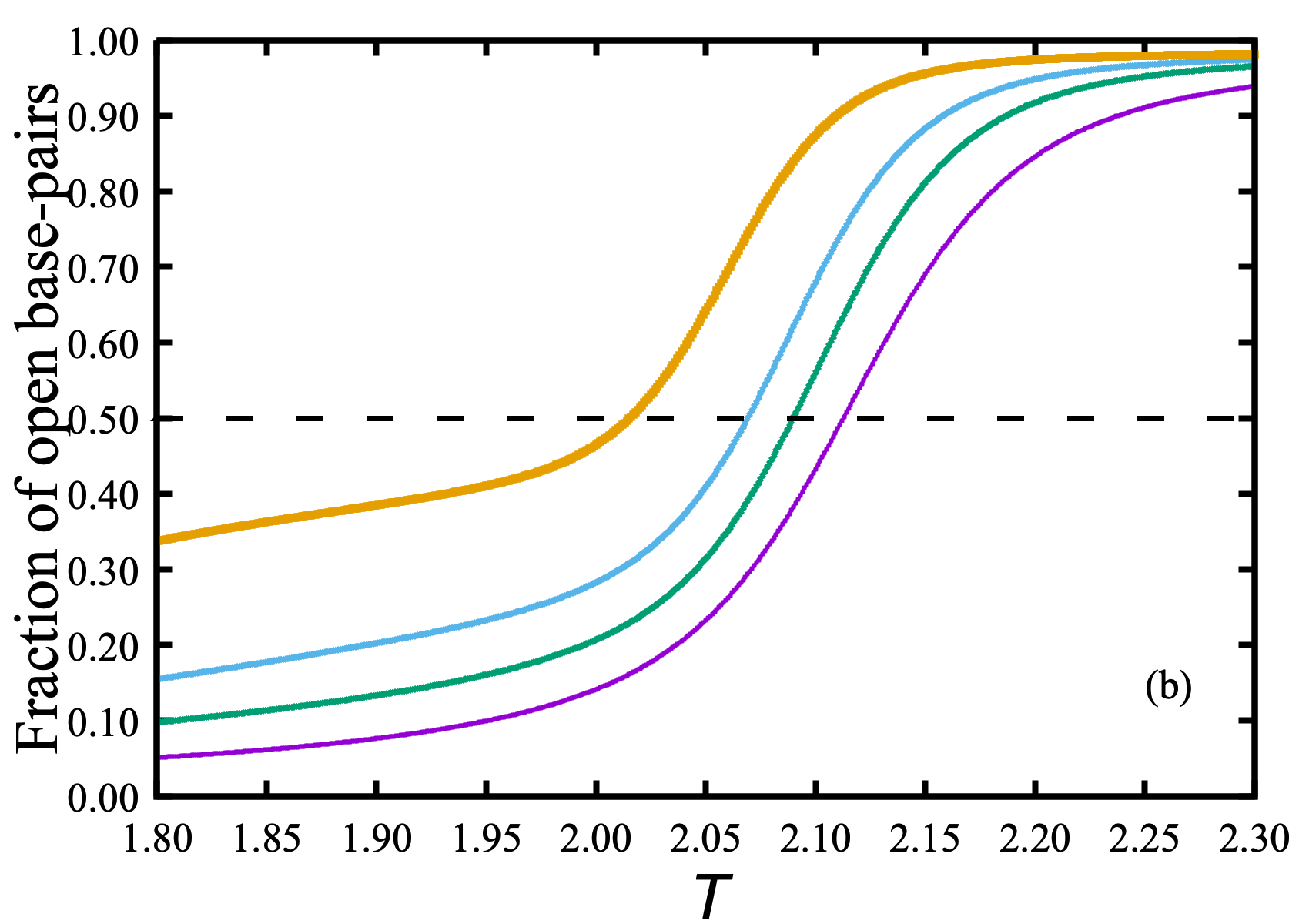}}
  \caption{Variation of fraction of open base-pairs with temperature at $f=0$ using (a) BD simulation (b) GNM. It shows that the effect of loop is similar in the two models.}
  \label{loop entropy}
\end{figure}

\end{document}